\documentclass[numberedappendix, twocolumn, apj]{emulateapj}

\usepackage{latexsym}
\usepackage{graphicx}
\usepackage{amssymb,amsmath}
\usepackage{amsfonts}   
\usepackage{natbib}
\usepackage{times}
\begin{document}

\newcommand{\ww}{0.48\linewidth}
\newcommand{\valf}{v_\text{Alf}}
\newcommand{\params}{\Theta}
\newcommand{\data}{{\bf D}}
\newcommand{\be}{\begin{equation}}
\newcommand{\ee}{\end{equation}}
\newcommand{\like}{{\mathcal L}}
\newcommand{\rt}[1]{{\bf RT: #1}}
\newcommand{\prof}{{\mathfrak L}}
\newcommand{\chisq}{\chi^2}
\newcommand{\mdl}{{\mathcal{M}}}
\newcommand{\norm}[2]{${\mathcal N}(#1, #2)$}

\newcommand{\icrc}{Int.\ Cosmic Ray Conf.}

\newcommand{\galprop}{GALPROP}
\newcommand{\Berat}{$^{10}$Be/$^9$Be}
\newcommand{\gray}{$\gamma$-ray}
\def\Dpp{D_{pp}}
\def\Dxx{D_{xx}}
\def\ddp{\frac{\partial}{\partial p}}
\def\dfdt{\frac{\partial f(\vec p)}{\partial t}}
\def\dfdp{\frac{\partial f(p)}{\partial p}}
\def\dNpdt{\frac{\partial \psi}{\partial t}}
\def\dNpdp{\frac{\partial \psi}{\partial p}}
\def\Xco{$X_{\rm CO}$}
\newcommand{\hi}{H~{\sc i}}
\newcommand{\hii}{H~{\sc ii}}
\newcommand{\range}[2]{$[#1,#2]$} 

\title{Constraints on cosmic-ray propagation models from a global Bayesian analysis}
\shorttitle{Constraints on cosmic ray}
\shortauthors{Trotta et al.}



\author{R. Trotta\altaffilmark{1}}
\altaffiltext{1}{Astrophysics Group, Imperial College London,
	Blackett Laboratory, Prince Consort Road, London SW7 2AZ, UK }

\author{G. J\'{o}hannesson\altaffilmark{2}} 
\altaffiltext{2}{Science Institute, University of Iceland, Dunhaga 3, IS-107 Reykjavik, Iceland}

\author{I.~V. Moskalenko\altaffilmark{3,4}} 
\altaffiltext{3}
{Hansen Experimental Physics Laboratory, Stanford University, Stanford, CA 94305
}\altaffiltext{4}
{Kavli Institute for Particle Astrophysics and Cosmology, Stanford University, Stanford, CA 94305}



\author{T.~A. Porter\altaffilmark{3}}   

\author{R. Ruiz de Austri\altaffilmark{5}}
\altaffiltext{5}{Instituto de F\'isica Corpuscular, IFIC-UV/CSIC, Valencia, Spain}

\author{A.~W. Strong\altaffilmark{6}} 
\altaffiltext{6}{Max-Planck-Institut f\"ur extraterrestrische Physik, Postfach 1312, D-85741 Garching, Germany}

\begin{abstract}
Research in many areas of modern physics such as, e.g., indirect 
searches for dark matter and particle acceleration
in SNR shocks, rely heavily on studies of cosmic rays (CRs) and 
associated diffuse emissions (radio, microwave, X-rays, \gray{s}).
While very detailed numerical models of CR propagation exist, a quantitative 
statistical analysis of such models has been so far hampered by the 
large computational effort that those models require. Although statistical 
analyses have been carried out before using semi-analytical 
models (where the computation is much faster), the evaluation of the 
results obtained from such models is difficult, as they necessarily 
suffer from many simplifying assumptions, 
The main objective of this paper is to present a working method for  
a full Bayesian parameter estimation for a 
\emph{numerical} CR propagation model. For this study, we use the 
\galprop{} code, the most advanced of its kind, that uses
astrophysical information, nuclear and particle data as 
input to self-consistently 
predict CRs, \gray{s}, synchrotron and other observables.
We demonstrate that a full Bayesian analysis is possible using 
nested sampling and Markov Chain Monte Carlo methods (implemented in 
the SuperBayeS code) despite the heavy 
computational demands of a numerical propagation code.
The best-fit values of parameters found in this analysis are in agreement 
with previous, significantly simpler, studies also based on \galprop.
\end{abstract}


\maketitle

\section{Introduction}

A large number of outstanding problems in physics and astrophysics are 
connected with studies of CRs and the diffuse emissions 
(radio, microwave, X-rays, \gray{s}) produced during their propagation in 
interstellar space.
These include: indirect searches for dark matter, the origin and 
propagation of CR; particle acceleration in putative CR sources -- such as 
supernova remnants (SNRs) -- and the interstellar medium (ISM); CR in other 
galaxies and the role they play in galactic evolution; studies of our local 
Galactic environment; CR propagation in the heliosphere; and the origin of 
extragalactic diffuse emissions.

The involved nature of these studies 
requires \emph{reliable and detailed calculations}.
Our current knowledge of CR propagation in the Galaxy is based on a 
large body of observational data together with substantial theoretical 
background: 
the latest developments in CR acceleration and transport mechanisms, 
detailed maps of the three-dimensional Galactic gas distribution, 
detailed studies of the interstellar dust, radiation field, 
and magnetic field, as well as 
up-to-date particle and nuclear cross section data and codes.
However, 
the number of parameters in realistic models incorporating all of 
this information is large, and using the available data to 
perform statistical inference on the models' free 
parameters is a highly non-trivial task. 
So far, this has only been possible with analytical or semi-analytical 
models where the computation is 
fast \citep[e.g.,][]{Donato2002,Maurin2001,Maurin2002,Maurin2010,Putze2010}.
But, such models necessarily require many simplified assumptions to allow 
the problem to be analytically tractable
and to reduce the computational load, making the estimation of 
the confidence level of their results difficult.
More realistic treatments using the analytic approach 
lead to a growing complexity of the formulae, thus removing any 
computational advantage over the purely 
numerical approach \citep[see, e.g.,][]{SMP2007}.

The \galprop{}\footnote{http://galprop.stanford.edu} 
code is the most advanced of its kind.
\galprop{} uses 
astronomical information and other data as input 
to self-consistently predict CRs, \gray{s}, synchrotron and other 
observables.
The code provides a full numerical calculation of the CR spectra and 
intensities, together with the diffuse emissions associated with the CRs 
interacting with the interstellar gas, radiation, and magnetic fields.
In this paper we introduce the methodology for a  
complete, fully numerical inference for propagation models parameters, 
using 
representative CR
data in a Bayesian statistical framework. 
We give results from a global analysis of CR isotope 
data, obtained by using \galprop{} to predict CR spectra and 
a modified version of the SuperBayeS code\footnote{http://superbayes.org} 
to carry out the statistical analysis. 
We demonstrate that improvements to the \galprop{} code, including 
parallelization, coupled with highly efficient sampling techniques 
and Bayesian methods now allow a fully numerical exploration of the 
parameter space of the most realistic models 
incorporating CRs, \gray{s}, etc., as well as experimental and 
theoretical uncertainties. 

The fully Bayesian approach to the problem of deriving constraints for CR propagation models parameters has several advantages.
Firstly, the higher efficiency of Bayesian methods allows us to carry out a global statistical analysis of the whole parameter space, rather than be limited to scanning a reduced number of dimensions at the time. This is important in order 
to be able to fit simultaneously all relevant CR parameters and to explore degeneracies.
Secondly, we can marginalize (i.e., integrate over) the parameters one is not interested in at almost no additional computational costs, thus obtaining probability distributions for the parameters of interest that fully account for correlations in the global parameter space. Thirdly, our method returns not only a global best fit point, but also statistically well-defined errors on the parameters, which is one of the most important achievements of this work. Finally, we are able to include in our analysis a large number of ``nuisance'' parameters (such as modulation potentials and experimental error rescaling parameters, see below for details) that mitigate the impact of potential systematic errors in the data and/or in the theoretical model, thus making our fits much more robust. Bayesian inference however requires to choose priors for the parameters involved. This is done very carefully in the present work, and we demonstrate below that our results do not depend strongly on the choice of priors, which again is an hallmark of a robust statistical analysis. 

\section{Cosmic-ray Propagation in the Galaxy}\label{CR_propagation}

Here we provide a brief overview of CR production and propagation, 
more information can be found in a recent review by \citet{SMP2007}.

The sources of CRs are believed to be supernovae (SNe) and SNRs, superbubbles,
pulsars, compact objects in close binary systems, and stellar winds.
Observations of X-ray and \gray{} emission from these objects reveal
the presence of energetic particles, thus testifying to efficient
acceleration processes in their neighborhood.  
Particles accelerated near the sources
propagate tens of millions years in the interstellar medium (ISM) while
their initial spectra and composition change.
The destruction of primary nuclei via spallation gives rise to secondary 
nuclei and isotopes that are rare in nature, antiprotons, and charged and 
neutral pions that decay producing secondary positrons, electrons, and \gray{s}.

Modeling CR propagation in the ISM includes the solution of the partial 
differential 
equation describing the transport 
with a given source distribution and boundary conditions for all CR species.
The diffusion-convection equation, sometimes incorporating diffusive 
reacceleration in the ISM, 
is used for the transport process and has proven to be remarkably successful 
despite its relative simplicity. 
For CR nuclei, relevant processes during propagation include nuclear 
spallation, 
production of secondary particles, radioactive decay, electron K-capture 
and stripping, 
with the energy losses due to ionization and Coulomb scattering.
For the propagation of CR electrons and positrons, spallation, 
radioactive decay, etc., are not
relevant, while the energy losses are due to ionization, 
Coulomb scattering, bremsstrahlung (with
the neutral and ionized gas), inverse Compton (IC) scattering, and 
synchrotron emission.

Measurements of stable and radioactive secondary CR nuclei yield 
the basic information necessary to probe large-scale Galactic
properties, such as the diffusion coefficient and halo size, the
Alfv\'en  velocity and/or the convection velocity, as well as
the mechanisms and sites of CR acceleration.
Knowing the number density of primary nuclei from satellite and
balloon observations, the production cross-sections from 
accelerator experiments, and the gas distribution from astronomical
observations, the production rate of secondary nuclei can be 
calculated within the context of a given propagation model.
Stable secondary
CR nuclei (e.g., $_{5}$B) can be used to determine ratio of halo size to the 
diffusion coefficient, while
the observed abundance of radioactive CR isotopes 
($^{10}_{4}$Be, $^{26}_{13}$Al, $^{36}_{17}$Cl, $^{54}_{25}$Mn) allows
the separate determination of halo size and diffusion coefficient
\citep[e.g.,][]{Ptuskin1998,SM1998,Webber1998,MMS2001}.
However, the interpretation of the sharp peaks observed in the 
secondary to primary CR
nuclei ratios (e.g., $_5$B/$_6$C, [$_{21}$Sc+$_{22}$Ti+$_{23}$V]/$_{26}$Fe) 
at relatively low energies, $\sim 1$-few GeV/nucleon, is model-dependent.

The solar modulation of the CRs during their propagation in the
heliosphere significantly modifies the interstellar spectra 
below $\sim 20$ GeV/nucleon.
The modulated spectra
are the actual ones measured by balloon-borne and spacecraft instruments.
Modulation models are based on the solution of the \citet{Parker1965} equation
\citep[e.g., see reviews by][]{Potgieter1998,Heber2006}.
The particle transport to the inner heliosphere is mainly determined
by spatial diffusion, convection with the solar wind, drifts, 
and adiabatic cooling.
Realistic time-dependent three-dimensional hydrodynamic models incorporating
these effects have been developed \citep[e.g.,][]{Florinski2003,Langner2006,Potgieter2004}.
The often-used method of \cite{GleesonAxford1968}, the 
so-called ``force-field''
approximation, employs a single 
parameter -- the ``modulation potential'' -- 
to characterize the strength of the modulation effect on the CR spectra 
as it varies over the solar cycle.
The force-field approximation has no predictive power 
as the modulation potential depends on the assumed interstellar spectrum of 
CR species.
However, it can be a useful low-energy parameterization
for a given interstellar spectrum. 

Closely connected with the CR propagation is the production of the
Galactic diffuse \gray{} emission.
This is comprised of three
components: $\pi^0$-decay, bremsstrahlung, and IC.
Cosmic-ray nuclei interacting inelastically with the 
interstellar gas produce $\pi^0$s that decay to \gray{s}.
The CR electrons and positrons interact with the interstellar gas
and produce bremsstrahlung, and with the interstellar radiation field (ISRF) 
via IC scattering producing \gray{s}.
Since the \gray{s} are undeflected by magnetic fields and 
absorption in the ISM is 
negligible \citep{MPS2006}, they provide the information necessary 
to directly probe CR spectra and intensities in distant 
locations \citep[see][for a review]{MSR2004}.
However, the interpretation of such observations is complicated 
since the observed \gray{} 
intensities are the line-of-sight integral of a sum of the three 
components of the diffuse
Galactic \gray{} emission, an isotropic 
component (often described as ``extragalactic'', but 
this not completely certain),
unresolved sources, together with instrumental background(s). 
Proper modeling of the diffuse \gray{} emission, including the 
disentanglement of the different components, requires well developed 
models for the ISRF and gas densities, together with the 
CR propagation \citep[see, e.g.,][]{SMR2000,SMR2004}. 
For recent measurements of the diffuse \gray{} emission by the \emph{Fermi} 
Large Area Telescope (LAT), see \citet{Abdo2009midlatitudes,Abdo2009localdiffuse,Abdo2010egb}.
Global CR-related properties of the Milky Way galaxy are 
calculated in \citet{Strong2010}.

\section{\galprop{} code}\label{galprop}

The \galprop{} project began in late 1996 and
has now 15 years of development behind it. 
The key concept underlying the \galprop{} code is that various 
kinds of data, e.g., 
direct CR measurements including primary and secondary nuclei, 
electrons and positrons, 
\gray{s}, synchrotron radiation, and so forth, are all related to the 
same astrophysical 
components of the Galaxy and hence have to be modeled self-consistently.
The code, originally written in FORTRAN90, was made public in 1998. 
A version rewritten in C++ was produced in 2001, and the most recent public 
version 54 was recently released \citep{GalpropWebrun}.
The code is available 
from the dedicated website where a facility for users to run the 
code via online forms in a web-browser is also 
provided\footnote{http://galprop.stanford.edu/webrun}.

The \galprop{} code solves the CR transport equation with a given source 
distribution and boundary conditions for all CR species. 
This includes a galactic wind (convection), diffusive reacceleration in the
ISM, energy losses, nuclear fragmentation, 
radioactive decay, and production of secondary particles and isotopes:

\begin{eqnarray}
\label{eq.1}
\frac{\partial \psi}{\partial t} 
&=& q({\mathbf r}, p)+ \nabla \cdot ( \Dxx\nabla\psi - {\mathbf V}\psi )
+ \ddp\, p^2 \Dpp \ddp\, \frac{1}{p^2}\, \psi \nonumber\\
&-& \frac{\partial}{\partial p} \left[\dot{p} \psi
- \frac{p}{3} \, (\nabla \cdot {\mathbf V} )\psi\right]
- \frac{1}{\tau_f}\psi - \frac{1}{\tau_r}\psi\, ,
\end{eqnarray}

\noindent
where $\psi=\psi ({\mathbf r},p,t)$ is the density per unit of total
particle momentum, $\psi(p)dp = 4\pi p^2 f({\mathbf p})$ in terms of
phase-space density $f({\mathbf p})$, $q({\mathbf r}, p)$ is the source term,
$\Dxx$ is the spatial diffusion coefficient, ${\mathbf V}$ is the
convection velocity, reacceleration is described as diffusion in
momentum space and is determined by the coefficient $\Dpp$,
$\dot{p}\equiv dp/dt$ is the momentum loss rate, $\tau_f$ is the time
scale for fragmentation, and $\tau_r$ is the time scale for
radioactive decay. 
The numerical solution of the transport equation is
based on a Crank-Nicholson \citep{Press1992} implicit second-order
scheme. 
The spatial boundary conditions assume free particle escape, e.g.,  
$\psi(R_h,z,p) = \psi(R,\pm z_h,p) = 0$,
where $R_h$ and $z_h$ are the boundaries for a cylindrically symmetric geometry.

Since the grid involves a 3D $(R,z,p)$ or 4D $(x,y,z,p)$ problem
(spatial variables plus momentum)
``operator splitting'' is used to handle the implicit solution.
For a given halo size the diffusion coefficient, as a function of
momentum and the reacceleration or convection parameters, is determined
from secondary/primary ratios. 
If reacceleration is included, the momentum-space diffusion
coefficient $D_{pp}$ is related to the spatial coefficient $\Dxx$ 
($= \beta D_0\rho^{\delta}$) \citep{Berezinskii1990,Seo1994}:

\begin{equation}
\label{eq.2}
\Dpp\Dxx = {4 p^2 \valf^2\over 3\delta(4-\delta^2)(4-\delta) w}\ ,
\end{equation}
where $w$ characterizes the level of turbulence (we take $w = 1$ since only the quantity
$\valf^2 /w$ is relevant),
and $\delta=1/3$ for a Kolmogorov spectrum
of interstellar turbulence or $\delta=1/2$ for a Kraichnan 
cascade (but can also be arbitrary), $\rho \equiv pc/Ze$ is the magnetic 
rigidity where $p$ is momentum and $Ze$ is the charge, $D_0$ is a 
constant, and $\beta= v/c$.
Non-linear wave damping \citep{Ptuskin2006} can also be included 
via specifying parameters in the configuration {\it galdef} file.

Production of secondary positrons and electrons is calculated as 
described in \citet{MS1998} with a correction by \citet{Kelner2006}.
Secondary pion production is calculated using the formalism by 
\citet{Dermer1986a,Dermer1986b}, which combines isobaric \citep{Stecker1970}
and scaling \citep{Badhwar1977,Stephens1981} models of the reaction, 
as described in \citet{MS1998}, or 
using a parameterization developed by \citet{Kamae2005}.
Bremsstrahlung is calculated as described in \citet{SMR2000}.
The IC scattering is treated using the appropriate formalism for an
anisotropic radiation field developed by \citet{MS2000}
using the full spatial and angular distribution of the ISRF calculated using
the \emph{FRaNKIE} code \citep{PS2005,Porter2008}. 

The distribution of Galactic CR sources is based on \emph{Fermi}-LAT \gray{} data.
For this study, we use $f_{CR}(R) = (R/R_0)^\alpha e^{-\beta(R-R_0)}$, i.e., 
normalized to 1 at $R=R_0$, where 
$\alpha=1.25$, and $\beta= 3.56$.
The profile is flattened for the outer Galaxy compared to earlier 
parameterizations used, as 
suggested from recent Fermi studies of the 2nd Galactic 
quadrant \citep{Tibaldo2009}. 

The \gray{s} are calculated using the propagated CR distributions, 
including a contribution from secondary
particles such as positrons and electrons from inelastic processes in the ISM 
that increases the \gray{} flux at MeV energies \citep{SMR2004,Porter2008}.
Gas-related \gray{} intensities ($\pi^0$-decay, bremsstrahlung) 
are computed from the emissivities as a
function of $(R,z,E_\gamma)$ using the column densities of \hi\ and
H$_2$ for galactocentric annuli based on recent 21-cm and CO survey data with 
a more accurate assignment of the gas to the Galactocentric 
rings than earlier versions.
The synchrotron emission is computed using the 
Galactic magnetic field model that can be chosen from among various 
models taken from the 
literature, suitably parameterized to allow fitting to the observations.
The line-of-sight integration of the corresponding emissivities with the 
distributions of gas, ISRF,
and magnetic field yields \gray{} and synchrotron skymaps.
Spectra of all species on the chosen grid and the \gray{}
and synchrotron skymaps are output in standard astronomical 
formats for direct comparison
with data, e.g., 
HEALPix\footnote{http://healpix.jpl.nasa.gov} \citep{Gorski2005}, 
\emph{Fermi}-LAT MapCube format for use with LAT Science Tools 
software\footnote{http://fermi.gsfc.nasa.gov/ssc/data/analysis}, etc.

Cross-sections are based on the extensive LANL database,
nuclear codes, and parameterizations
\citep{Mashnik2004}. 
The most important isotopic production cross-sections
(2H, 3H, 3He, Li, Be, B, Al, Cl, Sc, Ti, V, and Mn)
are calculated using our fits to major production channels
\citep{MM2003,Moskalenko2003}.
Other cross-sections are calculated using
phenomenological approximations by \citet{Webber2003} 
and/or \citet{Silberberg1998} renormalized to
the data where they exist. 
The nuclear reaction network is built using the
Nuclear Data Sheets.

The \galprop{} code
computes a complete network of primary, secondary and tertiary CRs
production starting from input source abundances.
Starting with the heaviest primary nucleus considered (e.g.\
$^{64}$Ni) the propagation solution is used to compute the source term
for its spallation products $A-1$, $A-2$ and so forth, which are 
then propagated in turn, and so
on down to protons, secondary electrons and positrons, and
antiprotons.  
To account for some special $\beta^-$-decay cases (e.g.,
$^{10}$Be$\to^{10}$B) the whole loop is repeated twice. 
\galprop{} includes K-capture and electron stripping processes as 
well as knock-on electrons.
The inelastically scattered protons and antiprotons are treated 
as separate components (secondary protons, tertiary antiprotons).
In this way secondaries, tertiaries, etc., are included.
Primary electrons are treated separately.  

Further details on improvements to the code, including parallelization 
and other optimizations,
improvements in line-of-sight integration routines, and so forth, 
can be found at the aforementioned website.

\section{Methodology} 
\label{methodology}


\subsection{Bayesian Inference}

The goal of this paper is to determine constraints on the propagation 
model parameters (introduced below) from observed CR spectra 
and we adopt a Bayesian approach to parameter 
inference \citep[see e.g.][for further details]{Trotta2008}. 
Bayesian inference is based on the posterior probability 
distribution function (pdf) for the parameters, 
which updates our state of knowledge from the prior by taking 
into account the information contained in the likelihood. 
A recent application to CR propagation models is 
given in \citet{Maurin2010} and \citet{Putze2010}. 
Denoting by $\params$ the vector of parameters one is interested in 
constraining, and by $\data$ the 
available observations, Bayes Theorem reads

\be \label{eq:bayes}
P(\params|\data) = \frac{P(\data | \params)P(\params)}{P(\data)},
\ee 

\noindent
where $P(\params|\data)$ is the posterior distribution on the 
parameters (after the observations 
have been taken into account), $P(\data | \params) = \like(\params)$ is 
the likelihood function (when 
considered as a function of $\params$ for the observed data $\data$) 
and $P(\params)$ is the prior 
distribution, which encompasses our state of knowledge about the 
value of the parameters before we have seen the data. 
Finally, the quantity in the denominator of eq.~\eqref{eq:bayes} is the 
Bayesian 
evidence (or model likelihood), a normalizing constant that does not 
depend on $\params$ and can be 
neglected when interested in parameter inference. 
The evidence is obtained by computing the average of the likelihood under the 
prior (so that the r.h.s.\ of eq.~[\ref{eq:bayes}] is properly normalized), 

\be \label{eq:evidence}
P(\data) = \int  P(\data | \params) P(\params) d\params . 
\ee

\noindent
The evidence is the prime quantity for Bayesian model comparison, which 
aims at establishing 
which of the available models is the ``best'' one, i.e., the one that 
fits the data best while 
being the most economical in terms of parameters, thus giving a quantitative 
implementation of Occam's razor  \citep[see, e.g.,][]{Trotta2007}. 
The evidence can also be used to assess the constraining power of 
the data \citep{Trotta2008a} and 
to carry out consistency checks between observables \citep{Feroz2009}. 

Together with the model, the priors for the parameters 
which enter Bayes' theorem, eq.~\eqref{eq:bayes}, must be specified. 
Priors should summarize our
state of knowledge and/or our theoretical prejudice about the parameters 
before we 
consider the new data, and for the parameter inference step the prior 
for a new observation 
might be taken to be the posterior from a previous 
measurement \citep[for model comparison 
issues the prior is better understood in a different way, see][]{Trotta2008}. 

The problem is then fully specified once we give the likelihood 
function for the 
observations (see section~\ref{sec:like} below). 
The posterior distribution $P(\params|\data)$ is determined numerically 
by drawing samples from it and 
Markov Chain Monte Carlo (MCMC) techniques can be used for this 
purpose. 
In this paper we use both Metropolis-Hastings MCMC and the 
MultiNest algorithm, which implements nested sampling and provides a 
higher efficiency, guarantees a 
better exploration of degeneracies and multimodal posteriors, 
and computes the Bayesian 
evidence as well (which is difficult to extract from MCMC methods). 

\subsection{Propagation Model Parameters}\label{parameters}

As a test case for this study we choose the diffusion-reacceleration model, 
which has been used in a 
number of studies utilizing the GALPROP code \citep[e.g.,][and references therein]{Moskalenko2002,SMR2004,Ptuskin2006,Abdo2009midlatitudes}. The 
source distribution is specified in Section~\ref{galprop}.

In this model the spatial diffusion coefficient is given by

\begin{equation}
D_{xx} = \beta D_{0} \left( \frac{\rho}{\rho_0}\right)^\delta
\label{Dxx}
\end{equation}

\noindent
where $D_{0}$ is a free normalization at the fixed 
rigidity $\rho_0 = 4\times 10^3$ MV. 
The power-law index is
$\delta=1/3$ for Kolmogorov diffusion (see Section \ref{galprop}), but we 
take it as a free parameter for the purposes of this study. 
Fitting the B/C ratio below 1 GeV in reacceleration models is known to 
require large values of $\valf$. 
In these models, a break in the injection spectra is required to 
compensate for the large bump in the propagated spectra at 
low energies/nucleon. 
Therefore, the CR injection spectrum is modeled as a broken power-law, with 
index below ($-\nu_1$) and above ($-\nu_2$) the break as free parameters, but 
with the location of the 
break
fixed at a rigidity $10^4$ MV. 
The other free model parameters are $\valf$, the halo size $z_h$,
and the normalization of the propagated CR proton 
spectrum at 100~GeV $N_p$, 
for a total of 7 free model parameters, as summarized in Table~\ref{tab:params}.
Other models discussed in the literature may be able to 
reproduce the B/C ratio without a 
break in the injection spectra, but the present paper is mainly intended 
as a presentation of the method and we defer a comprehensive study of 
different possibilities to a forthcoming paper.

The nuclear chain used starts at $^{28}$Si and proceeds down to protons.
The source abundances of nuclei $6\ge Z\ge 14$ have an important 
influence on the B/C and $^{10}$Be/$^9$Be ratios used in this study. 
In our analysis, they are fixed at values determined for ACE data at 
a few 100 MeV/nucleon \citep{Moskalenko2008}, but the values are 
assumed to hold also at the GALPROP normalization energy of 100 GeV/nucleon. 
The adopted relative source abundances of the most abundant 
isotopes (for particle flux in cm$^{-2}$ s$^{-1}$ (MeV/nucleon)$^{-1}$) are:
$^{4}$He: $7.199\times 10^4$ ,
$^{12}$C: 2819,
$^{14}$N: 182.8,
$^{16}$O: 3822,
$^{20}$Ne: 312.5,
$^{22}$Ne: 100.1,
$^{23}$Na: 22.84,
$^{24}$Mg: 658.1,
$^{25}$Mg: 82.5,
$^{26}$Mg: 104.7,
$^{27}$Al: 76.42,
$^{28}$Si: 725.7.
These values are relative to the proton normalization $N_p$ for a 
proton source abundance $1.06\times 10^6$, but this is only formal 
since the antiprotons, secondary positrons and gamma rays were computed 
from an independent fit to proton and He data.
$N_p$ is used only to normalize C and O to the data, via the 
ratios given above (other data like N are not used explicitly).

Special handling is required to treat the solar modulation of the 
propagated CR spectra, for which we introduce an extra nuisance 
parameter for each of the data set we consider. 
The motivation and choice of the Gaussian priors, with mean and 
standard deviation as 
given in Table~\ref{tab:params}, is described in Section \ref{CR_data}. 
In addition, we also introduce a set of parameters $\tau$ designed to 
mitigate the possibility that the fit be dominated by undetected 
systematic errors in the data, as explained in detail in the next section. 
Overall, we thus fit a total of 16 free parameters, including 7 model 
parameter, 4 modulation parameters, and 5 observational variance 
rescaling factors. 

\begin{deluxetable*}{lccc}
\tabletypesize{\footnotesize}
\tablecaption{\label{tab:params}
Summary of input parameters and prior ranges}
\tablecolumns{4}
\tablewidth{0pt}
\tablehead{
Quantity & Symbol & Prior range  & Prior type} 
\startdata
%
\multicolumn{4}{l}{\sc Diffusion model parameters $\params$} \smallskip\\
\quad Diffusion coefficient\tablenotemark{a} ($10^{28}$ cm$^2$ s$^{-1}$) & $D_0$ & \range{1}{12}  & Uniform\\
\quad Rigidity power law index & $\delta$  & \range{0.1}{1.0} & Uniform\\ 
\quad Alfv\'en speed (km s$^{-1}$) & $\valf$  & \range{0}{50} & Uniform\\
\quad Diffusion zone height (kpc) & $z_h$  & \range{1.0}{10.0} & Uniform\\ 
\quad Nucleus injection index below $10^4$ MV & $\nu_1$ & \range{1.50}{2.20} & Uniform\\
\quad Nucleus injection index above $10^4$ MV & $\nu_2$ &  \range{2.05}{2.50}& Uniform \\
\quad Proton normalization ($10^{-9}$ cm$^2$ sr$^{-1}$s$^{-1}$MeV$^{-1}$) & $N_p$ & \range{2}{8} & Uniform \medskip\\
\multicolumn{4}{l}{\sc Experimental nuisance parameters} \smallskip\\
\quad Modulation parameters $\phi$ (MV) 	& 					& 				& Gaussian prior\tablenotemark{b} \\
\qquad HEAO-3 		& $m_\text{HEAO-3}$	&  \range{420}{780} 	& \hspace{0mm} \norm{600}{60}\\ 
\qquad ACE  	& $m_\text{ACE}$  		&  \range{175}{475} 	& \hspace{0mm} \norm{325}{50} \\ 
\qquad CREAM 	& $m_\text{CREAM}$	&  \range{420}{780}	& \hspace{0mm} \norm{600}{50}\\ 
\qquad ISOMAX 	& $m_\text{ISOMAX}$	&  \range{370}{490}	& \hspace{0mm} \norm{430}{20}\\ 
\qquad ATIC-2  	& $m_\text{ISOMAX}$ 	& 0 				& Fixed (no modulation)\\ 
\quad Variance rescaling parameters ($j=1,\dots,5$)   & $\log\tau_j$  &  \range{-1.5}{0.0}	& Uniform on $\log\tau_j$

\enddata
\tablenotetext{a}{At $\rho=4\times10^3$ MV.}
\tablenotetext{b}{We use the notation  $\mathcal{N}(\mu, \sigma)$ to indicate a Gaussian distribution of mean $\mu$ and standard deviation $\sigma$. }
\end{deluxetable*}

\pagebreak[4]
\subsection{Cosmic Ray Data and Modulation}\label{CR_data}




For demonstration of the method we use the most accurate CR data sets 
available preferably taken near solar minimum\footnote{Most of the data are obtained via the database maintained at http://www.mpe.mpg.de/$\sim$aws/propagate.html}.

The B/C ratio is well-measured by
a number of space- and balloon-borne missions. 
The HEAO-3 data  
\citep{Engelmann1990} remain the most accurate to date in the
energy range 0.6--35 GeV/nucleon and have been recently confirmed by  
PAMELA (R. Sparvoli, private comm.). 
At higher energies, from 30 GeV/nucleon -- 1 TeV/nucleon, we 
use ATIC-2 \citep{Panov2008} and CREAM-1 data \citep{Ahn2008}.
At low energies, the Voyagers 1 and 2 \citep{Lukasiak1999}, Ulysses  
\citep{Duvernois1996}, and ACE \citep{deNolfo2006} data agree with  
each other, while the ACE data (50--200 MeV/nucleon) have the smallest  
statistical error. 
Therefore, we use the ACE measurements corresponding to the solar minimum
conditions \citep{George2009}.

The $^{10}$Be/$^9$Be ratio is most accurately  
measured (70--145 MeV/nucleon) by ACE \citep{Yanasak2001}, which we include 
in our fit. 
Those measurements are in agreement 
with Voyagers 1 and 2  \citep{Lukasiak1999}, and Ulysses \citep{Connell1998} 
data. 
At  higher energies (per nucleon) there are only two data 
points by ISOMAX \citep{Hams2004} with very large error  
bars, which we however include in the fit. 

We also use the carbon and oxygen spectra as measured by 
ACE at the solar minimum \citep{George2009}
and by HEAO-3 \citep{Engelmann1990}.

As mentioned above, a very important issue is the treatment of the 
heliospheric modulation.
We fit to the CR data in the whole energy range from some 10 MeV/nucleon 
to TeV energies. 
However, a comparison of calculated CR spectra, the elemental and 
isotopic ratios 
with low-energy data (below $\sim$20 GeV/nucleon)
measured inside the heliosphere requires care as the calculated 
spectra depend significantly on the treatment of the heliospheric modulation. 
As mentioned in section~\ref{CR_propagation}, the modulation can be
realistically treated with full 3-dimensional models, but application of 
such models to the current study does not seem feasible 
since the number of free parameters and the computing requirements 
would considerably increase.
Currently, it is only possible to 
use a simple force-field approximation \citep{GleesonAxford1968}, which 
is characterized with the value of the modulation potential.
However, directly using the modulation potentials
from different experiments is problematic because they 
cannot be interpreted independently from the experiments themselves. 
The derived values depend on the choices of interstellar spectra used for
their analyses, which differ from experiment to experiment (and are sometimes
not provided).

To deal with this type of uncertainty, instead of fixing a 
collection of \emph{a priori} values
for the modulation potential, we allow some flexibility to the fits and 
include the modulation potentials as free nuisance parameters in our 
inference, with one free parameter per 
experiment (i.e, ACE, HEAO-3, ISOMAX and CREAM-1).
To avoid unphysical/implausible values, we adopt Gaussian priors with mean and 
standard deviation as given in Table~\ref{tab:params}, which are motivated 
by the estimated ballpark values of the modulation by the experimentalists.  
Notice that no modulation parameter is given for ATIC-2 as we only 
use high-energy data 
for that experiment and modulation is not relevant. 

\subsection{The Likelihood Function} \label{sec:like}

For a given set of the CR model parameters $\params$ and the 
modulation potential parameters $\phi$ 
(where $\phi = \{\phi_1, \dots, \phi_4\}$, with a different choice of 
the modulation potential for each data set employed) we can compute 
via \galprop{} the ensuing CR spectrum, as a function of 
energy, $\Phi_X(E, \params, \phi)$ for species $X$. 
Assuming Gaussian noise on the observations, we take the likelihood 
function for each observation of species $X$ at energy $E_i$ to be of the form 

\begin{eqnarray}
\label{eq:Gauss_like}
& P(\hat{\Phi}_X^{ij} |& \params, \phi) = \\
&& \frac{1}{\sqrt{2\pi}\sigma_{ij}}  \exp\left( -\frac{1}{2}\frac{\left(\Phi_X(E_i, \params, \phi) - \hat{\Phi}_X^{ij} \right)^2}{\sigma_{ij}^2} \right),
\nonumber
\end{eqnarray}

\noindent
where $\Phi_X(E_i, \params, \phi)$ is the prediction from the CR 
propagation model for species $X$ at energy $E_i$, $ \hat{\Phi}_X^{ij}$ is 
the measured spectrum, and $\sigma_{ij}$ is the reported standard deviation. 
The sub/superscript $i$ runs through the data points within each of the 
data sets $j$. 
We assume bins to be independent, so that the full likelihood function 
is given by the product of terms of the form given above:  

\be \label{eq:like1}
P(\data | \params,\phi) = \prod_{j=1}^{5}\prod_{i=1}^{N_j} P(\hat{\Phi}_X^{ij} | \params, \phi)
\ee 

\noindent
However, a careful analysis of a plot of the data points for each CR 
species reveals that there are fairly strong discrepancies between 
different data sets. 
This might point to either an underestimation of the actual experimental 
error bars or to undetected systematic errors between data sets. 
If some or all of the reported error bars are significantly 
underestimated, this would lead to a handful of data points incorrectly 
dominating the global fit, introducing systematic bias in the 
reconstructed value of the parameters. 
To mitigate against undetected systematics, we follow the procedure 
described in, e.g., \citet{Barnes2003}. 
For each data set we introduce in the likelihood a 
parameter $\tau_j$ ($j=1,\dots,5)$, whose function is to rescale 
the variance of the data points in order to account for possible 
systematic uncertainties. 
Therefore, eq.~\eqref{eq:Gauss_like} is modified:  

\begin{eqnarray} 
\label{eq:Gauss_like_rescaled}
& P(\hat{\Phi}_X^{ij} |& \params, \phi, \tau) = \\
&& \frac{\sqrt{\tau_j}}{\sqrt{2\pi}\sigma_{ij}} \exp\left( -\frac{1}{2}\frac{ \left(\Phi_X(E_i, \params, \phi) - \hat{\Phi}_X^{ij} \right)^2}{\sigma_{ij}^2/\tau_j} \right),
\nonumber
\end{eqnarray}

\noindent
The role of the set of parameters $\tau = \{ \tau_1, \dots, \tau_5 \}$, 
which we call ``error bar rescaling parameters'', is to allow for the 
possibility that the error bars reported by each of the experiments 
underestimate the true noise. 
We then add $\tau$ to the set of parameters $\params$ and sample 
over it, too, thus allowing the data themselves to decide whether 
there are systematic discrepancies in the reported 
error bars. 
A value $\tau_j < 1$ means that the data prefer a systematically 
larger value for the errors for data set $j$. 
Notice that $\tau_j$ not only appears in the exponential of the 
Gaussian in Eq.~\eqref{eq:Gauss_like_rescaled}, but also in the pre-factor, 
which, being proportional to $\sqrt{\tau_j}$, ensures that $\tau_j$ 
never attains a value arbitrarily close to 0 (implying infinite error bars). 
Furthermore, the variance scaling parameter $\tau$ also takes care of 
all aspects of the model that are not captured by the reported 
experimental error: this includes also theoretical 
errors (i.e., the model not being completely correct), errors in the 
cross section normalizations, etc. 

\subsection{Choice of Priors}

The full posterior distribution for the CR propagation 
model parameters $\params$, the variance rescaling 
parameters $\tau$ and the modulation parameters $\phi$ is given by
 
\be \label{eq:fullposterior}
P(\params, \phi, \tau,  | \data) \propto P(\data | \params, \phi, \tau) P(\params)P(\tau)P(\phi),
 \ee

\noindent
where the likelihood $P(\data | \params, \tau, \phi)$ is 
given by Eq.~\eqref{eq:like1} and \eqref{eq:Gauss_like_rescaled}. 
  
The priors $P(\params)$, $P(\phi)$ and $P(\tau)$ in 
Eq.~\eqref{eq:fullposterior} are specified as follows.
We take the prior on a set of model parameters, 
$P(\params)$, 
to be uniform on $\params$ with ranges as given in Table~\ref{tab:params}. 
As shown below, the posterior is reasonably well constrained 
and close to Gaussian for $\params$, hence we expect our 
results to be fairly independent of the prior choice.

This conclusion is strengthened by the inspection of the profile likelihood, which is obtained from our samples by maximizing the value of the likelihood along the hidden dimensions rather then integrating over the posterior. The profile likelihood statistics is independent of the priors, provided the parameter space has been sampled with sufficient resolution, and thus it constitutes a cross-check for the presence of large volume effects coming from the priors. Such volume effects are typically important when the priors play a major role in the inference, while they are usually negligible when the posterior is dominated by the likelihood \citep[in which case the prior influence is minimal, see e.g.][ for an illustration]{Trotta:2008bp}. We have found the profile likelihood to be in excellent agreement with the posterior pdf presented below, and therefore we do not consider it further in our results below. This means that the prior influence is small, and our results can be considered to be  robust with respect to reasonable changes in priors.

Regarding the modulation parameters, we adopt a Gaussian prior 
on each of them, informed by the values reported by 
each experiment (see Table~\ref{tab:params}), 
in order to avoid physically implausible values. 
The description of the experimental CR data sets can be 
found in Section \ref{CR_data}.
 
The $\tau_j$ are scaling parameters in the likelihood, and thus 
the appropriate prior is given by the Jeffreys' prior, which is uniform 
on $\log\tau_j$ (see \citealt{Barnes2003} or \citealt{Jaynes2003} 
for a justification). 
Therefore, we adopt the proper prior 

\be
P(\log\tau_j) = \left\{ 
\begin{array}{c l}
  2/3 \text{ for}  & \mbox{ for } -3/2 \leq \log\tau_j \leq 0  \\
  0 &\mbox{ otherwise}
\end{array}
\right.
\ee 

\noindent
that corresponds to a prior on $\tau_j$ of the form

\be
P(\tau_j) \propto \tau_j^{-1}.
\ee

The inclusion in our analysis of the nuisance parameters $\phi$ and $\tau$ (which are then marginalized over, see Eq.~[\ref{eq:marginalisation_continuous}]) has two main effects on our inference about the CR parameters of interest, $\params$. Firstly, it increases the robustness of our fit, since the nuisance parameters account for potential systematic effects in the data ($\tau$) and approximately capture the impact of solar modulation on the measurements ($\phi$). Secondly, it makes our CR constraints more conservative, since our marginalized errors on $\params$ fully account for all possible values of the nuisance parameters compatible with the data.  

\subsection{Sampling Algorithm}

A powerful and efficient alternative to classical MCMC methods has 
emerged in the last few years in the form of the 
so-called ``nested sampling'' algorithm, invented by 
John Skilling \citep{Skilling2004,Skilling2006,Feroz2008,Feroz2009}. 
Although the original motivation for nested sampling was to compute 
the evidence 
integral of eq.~\eqref{eq:evidence}, the recent development of the 
MultiNest algorithm \citep{Feroz2008,Feroz2009} has delivered an 
extremely powerful and versatile algorithm that has been demonstrated 
to be able to deal with
extremely complex likelihood surfaces in hundreds of dimensions
exhibiting multiple peaks. 

As samples from the posterior are
generated as a by-product of the evidence computation, nested
sampling can also be used to obtain parameter constraints in the
same run as computing the Bayesian evidence. 
In addition, multi-modal nested sampling exhibits an efficiency that is 
almost independent 
of the dimensionality of the parameter space being explored, thus beating 
the ``curse of dimensionality''.

The essential element of nested sampling is that the
multi-dimensional evidence integral is recast into a
1-dimensional integral. 
This is accomplished by defining the
prior volume $X$ as $d X \equiv P(\params)d
\params $ so that
 
\begin{equation} \label{eq:def_prior_volume}
  X(\lambda) = \int_{\like(\params)>\lambda} P(\params) d
  \params
\end{equation}

\noindent
where the integral is over the parameter space
enclosed by the iso-likelihood contour $\like(\params) =
\lambda$. 
So $X(\lambda)$ gives the volume of parameter space
above a certain level $\lambda$ of the likelihood. 
Then the
Bayesian evidence, Eq.~\eqref{eq:evidence}, can be written
as

\begin{equation} \label{eq:nested_integral}
  P(\data) = \int_0^1 \like(X) d X,
\end{equation}

\noindent
where $\like(X)$ is the inverse of
Eq.~\eqref{eq:def_prior_volume}. 
Samples from $\like(X)$ can be
obtained by drawing uniformly samples from the likelihood volume
within the iso-contour surface defined by $\lambda$. The
1-dimensional integral of Eq.~\eqref{eq:nested_integral} can be
obtained by simple quadrature, thus
 
\begin{equation}
  P(\data) \approx \sum_i \like(X_i) W_i ,
\end{equation}

\noindent
where the weights are $W_i = \frac{1}{2}(X_{i-1} - X_{i+1})$, 
\citep[see][for details]{Skilling2004,Skilling2006,Feroz2008,Feroz2009,Mukherjee2006}. 
It has been shown in the context of CMB data analysis in cosmology 
and in supersymmetry phenomenology studies that this technique 
reduces the number of 
likelihood evaluations by over an order of magnitude with respect to 
conventional MCMC. 

In this paper we adopt the publicly available 
MultiNest algorithm~\citep{Feroz2008}, as implemented in the 
SuperBayeS code \citep{Trotta2008a,Ruiz2006}, that we have interfaced 
with the \galprop{} code. 
First, we performed an exploratory scan with MultiNest, 
adopting 4000~live points in our 16-dimensional parameter space, with the 
aim of scouting the structure and degeneracy directions of the 
posterior distribution. 
An important chracteristic of MultiNest, which sets it apart from 
conventional MCMC methods, is its ability to sample reliably 
from multi-modal distributions. 
Therefore, it is highly desirable to employ MultiNest to perform 
scans of parameter spaces that 
have not been investigated before, as MultiNest will make it less 
likely to miss important substructure in the probability 
distribution when the latter is multi-modal. 

Our exploratory MultiNest scan gathered $\sim 10^5$ samples from 
the posterior, with an overall efficiency of about 10\% and a total 
computational effort of $\approx 13$ CPU years. 
This initial scan revealed a well behaved, unimodal distribution over the prior 
ranges given in Table~\ref{tab:params}. 
We then computed the 
parameter-set covariance matrix and adopted this as a Gaussian proposal 
distribution for a conventional Metropolis-Hastings MCMC scan. 
Since the proposal distribution was well matched to the posterior, our 
MCMC scan reached an efficiency of $\sim 15\%$, and we built 10 
parallel chains with 14000 samples each (after burn-in), for a 
total of $1.4\times10^5$ samples. 
We checked that the Gelman \& Rubin mixing criterion~\citep{Gelman92} is 
satisfied for all of our parameters (i.e., $R \ll 0.1$, where $R$ is the 
inter-chain variance divided by the intra-chain variance). 

We verified that the posterior distribution obtained with MCMC was in 
excellent agreement with the one obtained from MultiNest, which validates 
our results as the two sampling schemes are completely different. 
The results presented in this paper are obtained from the MCMC run, which 
allows for slightly smoother posterior distributions as it contains 40\% 
more samples than the MultiNest scan\footnote{The resulting chain of 
samples (including their statistical weight and likelihood) is provided 
as Supplementary Material, allowing the reader to make their own 
analysis, for example to investigate particular parameter correlations.}. 

In order to keep the computational cost within reasonable limits, we 
carry out our MultiNest and MCMC runs assuming a relatively coarse 
spatial and energy/nucleon grid for the CR propagation.
For each point in parameter space, we adopt $\delta z = 0.2$ kpc, 
$\delta R = 1$ kpc, and $\delta E = 1.4$. 
With these parameters, one likelihood evaluation takes approximately 
15~s on an 8-way 2.4~GHz Opteron CPU machine.  
We then reprocessed the MCMC samples using importance sampling, 
decreasing the spacing of the spatial grid by a factor of 2 in each 
direction while increasing the energy resolution to $\delta E = 1.1$. 
The computational cost per likelihood evaluation is increased by a 
factor of $\sim 20$, but allows a more precise computation of the CR spectra. 
The statistical distribution is adjusted accordingly, thus obtaining a 
posterior distribution that is close to what would have been obtained by 
running the scan at the higher resolution initially.
The advantage of using posterior sampling in this context is that the 
resampling of the points can be done in a massively 
parallel way (using 800~CPUs, the resampling step takes only a few hours). 
Although the best-fit point shifts somewhat after importance sampling, we 
have verified that the bulk of our probability distributions remain stable. 
We therefore conclude that the results presented here are robust with 
respect to increases in the spatial and energy resolution of the scan.

\subsection{Parameter Inference from Posterior Samples}

Once a sequence $\{ \params^{(0)},  \params^{(1)}, \dots,  \params^{(M)} \}$ of 
samples from the posterior pdf has been gathered, obtaining Monte Carlo
estimates of expectations for any function of the parameters is 
straightforward. 
For example, the posterior mean is given
by ($\langle \cdot \rangle$ denotes the expectation value
with respect to the posterior)

\begin{equation} \label{eq:expectation}
  \langle \params \rangle \approx \int  P(\params|\data)\params d\params
  = \frac{1}{M} \sum_{t=0}^{M-1} \params^{(t)},
\end{equation}

\noindent
where the second equality
follows because the samples $\params^{(t)}$ are generated from the
posterior by construction. 
In general, the expectation value of any function of the 
parameters $f(\params)$
is obtained as

\begin{equation} \label{eq:MC_estimate}
  \langle f(\params) \rangle \approx \frac{1}{M}\sum_{t=0}^{M-1}  f(\params^{(t)}).
\end{equation}

\noindent
It is useful to summarize the results of the
inference by giving the 1-dimensional {\em marginal probability}
for the $j^{\rm th}$ element of $\params$, $\params_j$, obtained by 
integrating out all other parameters from the posterior:
 
\begin{equation} \label{eq:marginalisation_continuous}
  P(\params_j|\data) = \int  P(\params|\data) d \params_1 \dots d
  \params_{j-1} d \params_{j+1} \dots d \params_{n},
\end{equation}

\noindent
where $ P(\params_j | \data)$ is the {\em marginal posterior} for the
parameter $\params_j$. 
From the posterior samples (obtained either by MCMC or MultiNest) it is 
straightforward to 
obtain and plot the marginal posterior on the left-hand-side of
Eq.~\eqref{eq:marginalisation_continuous}: since samples are drawn 
from the full posterior by construction, their density 
reflects the value of $P(\params|\data)$. 
It is then sufficient to divide the range of
$\params_j$ into a series of bins and {\em count the number of
samples falling within each bin}, ignoring the coordinates
values $\params_i$ (for $i\neq j$). 
The 2-dimensional posterior is defined in an analogous fashion. 
The 1-dimensional, 2-tail symmetric
$\alpha\%$ credible region is given by the interval (for the
parameter of interest) within $\alpha\%$ of where the samples are found, 
obtained in such a way that a fraction $(1-\alpha)/2$ of the samples are 
outside the interval on either side. 
In the case of a 1-tail upper (lower) limit, we report the value of the
quantity below (above) where $\alpha\%$ of the samples are found.

\begin{figure*}
\centerline{
\includegraphics[width=.7 \linewidth]{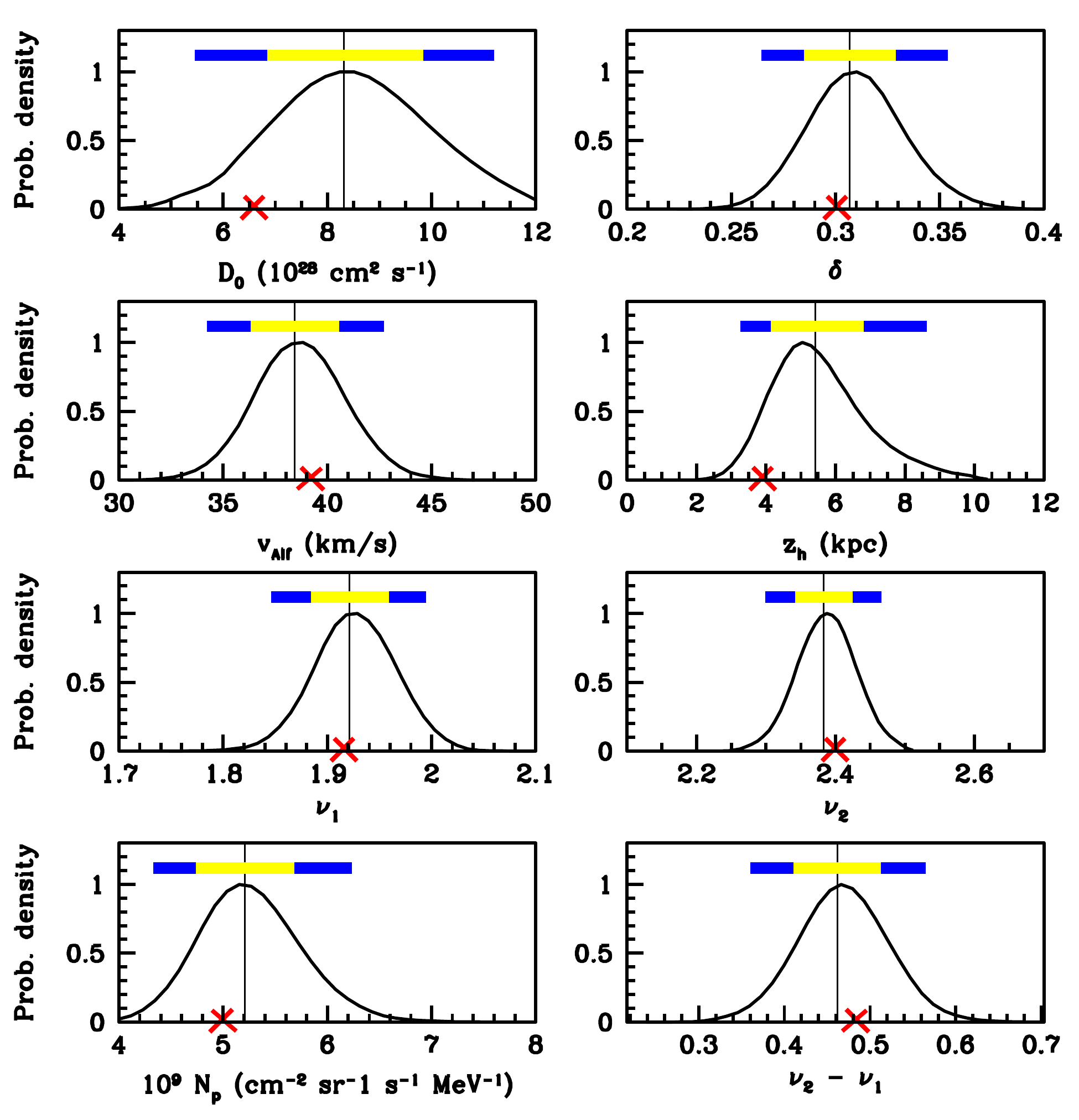}
}
\caption{1D marginalized posterior pdf normalized to the peak for the 
diffusion model parameters, with uniform priors assumed over the 
parameter ranges as in Table~\protect{\ref{tab:params}}. 
The red cross represents the best fit, the vertical thin line the 
posterior mean, and the horizontal bar the 68\% and 95\% error
ranges (yellow/blue, respectively). The bottom-right panel shows the pdf for the spectral index break.
 \label{1D_plots} }
\end{figure*}

\begin{deluxetable*}{lccc}[b!]
\tabletypesize{\footnotesize}
\tablecaption{\label{tab:params_constraints} 
Summary of constraints on all parameters
}
\tablecolumns{5}
\tablewidth{0pt}
\tablehead{
Quantity & Best fit & Posterior mean and & Posterior \\
&value & standard deviation & 95\% range} 
\startdata
%
\multicolumn{4}{l}{\sc Diffusion model parameters $\params$} \smallskip\\
\quad $D_{0} (10^{28}$ cm$^2$ s$^{-1}$) 	& 6.59 & $8.32\pm1.46$ & \range{5.45}{11.20} \\
\quad $\delta$  							& 0.30 & $0.31\pm 0.02$ &  \range{0.26}{0.35} \\ 
\quad $\valf$  (km s$^{-1}$)  				& 39.2 & $38.4 \pm 2.1$ & \range{34.2}{42.7}\\
\quad $z_h$ (kpc) 						& 3.9 & $5.4\pm1.4$ & \range{3.2}{8.6}\\
\quad $\nu_1$ 							& 1.91 & $1.92\pm0.04$ & \range{1.84}{2.00}  \\
\quad $\nu_2$ 							& 2.40  & $2.38 \pm 0.04$ & \range{2.29}{2.47}\\
\quad $N_p$ ($10^{-9}$ cm$^2$ sr$^{-1}$ s$^{-1}$ MeV$^{-1}$) & 5.00 & $5.20\pm0.48$ & \range{4.32}{6.23} \medskip\\
\multicolumn{4}{l}{\sc Experimental nuisance parameters} \smallskip\\
\quad Modulation parameters $\phi$ (MV)\\
\qquad HEAO-3	& 693 & $690\pm38$ & \range{613}{763}\\
\qquad ACE 	& 357 & $354\pm 22$ & \range{311}{398}\\
\qquad CREAM 	& 598 & $602\pm49$ & \range{503}{697}\\
\qquad ISOMAX 	& 416 & $430\pm20$ & \range{391}{470} \\
\qquad ATIC-2 	& 0 (fixed)& N/A & N/A  \smallskip\\

\multicolumn{4}{l}{\quad Variance rescaling parameters $\tau$}\\
\qquad HEAO-3  	& -0.60 & $-0.60 \pm 0.10$ & \range{-0.82}{-0.41} \\
\qquad ACE 	& -0.12 & N/A & $> -0.49$ (1-tail)\\
\qquad  CREAM 	&  0.00 & N/A & $>-0.53$ (1-tail)\\
\qquad ISOMAX & -0.21 & N/A & $>-1.21$ (1-tail)\\
\qquad  ATIC-2 	& -0.24 & N/A & $>-0.84$ (1-tail)

\enddata
\end{deluxetable*}

\section{Results} \label{results}

\subsection{Cosmic-Ray Propagation Model Parameter Constraints} 

\begin{figure*}[t!]
\centerline{
\includegraphics[width=.7\linewidth]{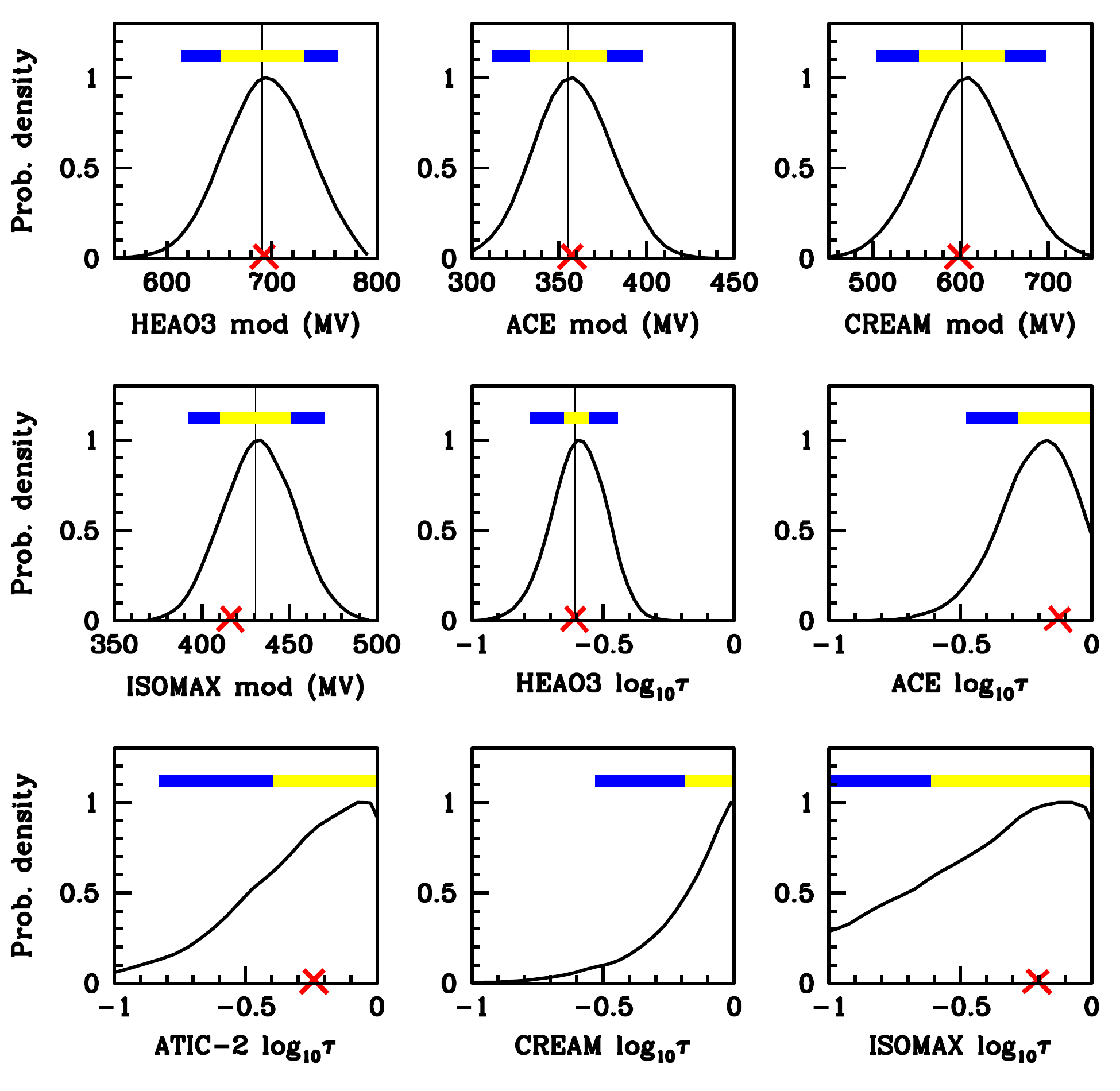}
}
\caption{As in Figure~\ref{1D_plots} but for the nuisance parameters 
used in the analysis. \label{1D_plots_nuisance} }
\end{figure*}

\begin{figure*}
\centerline{
\includegraphics[width=.7\linewidth]{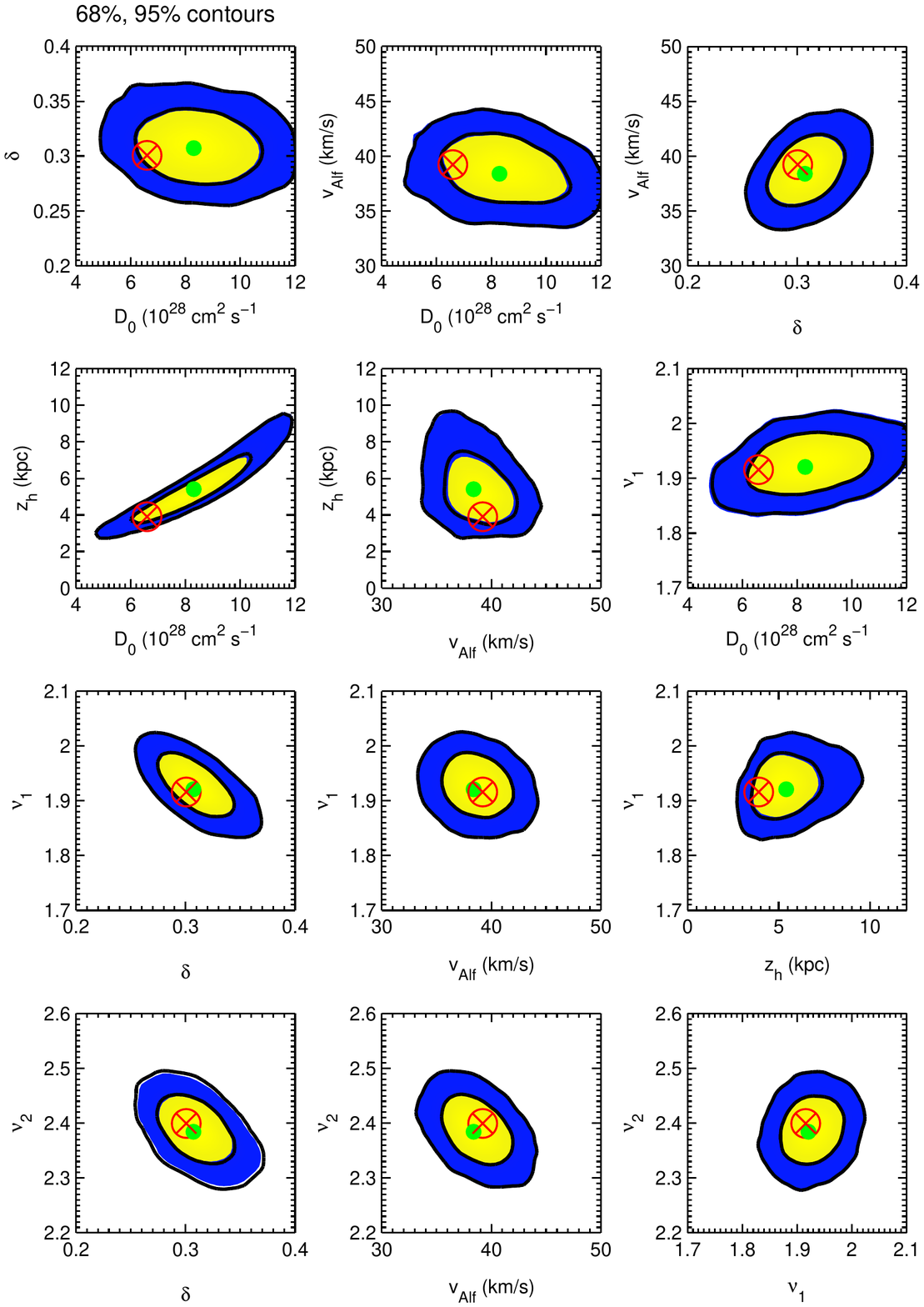}
}
\caption{2D marginalized posterior probability distributions for some 
parameter combinations. The yellow and blue regions enclose 68 and 95\% 
probability, respectively. 
The encircled red cross is the best fit, the filled green dot the 
posterior mean. 
\label{2D_plots}}
\end{figure*}

In Figures~\ref{1D_plots}, \ref{1D_plots_nuisance} we show 1-dimensional 
marginalized posterior 
probabilities for the propagation and nuisance parameters of the model.
The red cross represents the best fit, the vertical thin line the 
posterior mean, and the 
horizontal bar the 68\% and 95\% error ranges (yellow/blue, respectively). 
Two-dimensional constraints on some parameter combinations are 
presented in Figure~\ref{2D_plots}. 
The best-fit point and posterior ranges are 
summarized in Table~\ref{tab:params_constraints}. 
For reference the {\it galdef} parameter definition 
files with the best-fit parameter values presented in this study are 
available as Supplementary Material to this paper. 
These give precise definitions of the model used, which can be reproduced 
as required. 
The final MCMC chains from which Figures~\ref{1D_plots}-\ref{2D_plots} were 
produced are also available. 

We see that $\delta$ 
and $\valf$ are quite well constrained, with the posterior 
mean $\delta=0.31\pm 0.02$ being 
very close to the canonical value of $1/3$ for Kolmogorov diffusion. 
The Alfv\'en speed, $\valf = 38.4\pm2.1$ km s$^{-1}$, is higher than in earlier 
studies, but 
this is dictated by the fit to the 
ACE data on the B/C ratio at low energies.
The posterior intervals on the values of $D_{0}$ and $z_h$ are 
fairly large, $D_0 = (8.32 \pm 1.46) \times10^{28}$ cm$^2$ s$^{-1}$ and $z_h = 5.4 \pm 1.4$ kpc. 
The typical value of 4 kpc adopted in many studies is at the lower end 
of the viable range, but still within the 95\% 
interval, $z_h \in$ \range{3.2}{8.6} kpc. 
We can see from the $D_0$ vs.\ $z_h$ panel in Fig.~\ref{2D_plots} that the 
diffusion coefficient and the halo size are positively correlated, as 
expected. 

Other parameters exhibit less pronounced correlations. 
The injection indexes $\nu_1$ and $\nu_2$ are tightly constrained and 
almost uncorrelated 
(Figure~\ref{2D_plots}),
but 
this reflects the fact that the position of the injection spectral break is 
fixed in this analysis, so that the indices are determined 
by $\delta$ and $\valf$ with their narrow ranges.
The value of the injection index $\nu_2=2.38\pm 0.04$ provides a 
consistency check as the 
value of the sum $\nu_2+\delta$ should be close to the spectral 
indices of directly measured 
carbon and oxygen spectra $\sim$2.70, and indeed we find 
that  $\nu_2+\delta = 2.69 \pm 0.05$.
Comparison to the value of the injection index $\nu_1=1.92\pm 0.04$
shows that the spectral break required is $0.46\pm0.05$. The pdf for the spectral break is plotted in the bottom-right panel of Fig.~\ref{1D_plots}.
While at face value the break appears very statistically significant, 
it has be be kept in mind that the value found is dependent on the break energy, 
which was fixed in this analysis. Future analyses will allow more 
freedom in the form of the spectrum.

\subsection{Comparison with Our Previous Results}

In general, there is remarkable agreement between the ``by-eye'' fitting 
in the past~\citep[e.g.,][]{SM1998,SM2001a,Moskalenko2002,Ptuskin2006} 
and the parameter constraints  
found using the refined Bayesian inference analysis described in this paper. 
The posterior mean values of the diffusion 
coefficient $D_0=(8.32 \pm 1.46) \times10^{28}$ cm$^2$ s$^{-1}$ at 4 GV and the
Alfv\'en speed $\valf=38 .4\pm 2.1$ km s$^{-1}$ are also in fair agreement 
with earlier estimates  of $5.73\times10^{28}$ cm$^2$ s$^{-1}$ and 
36 km s$^{-1}$ \citep{Ptuskin2006}, respectively. 
The posterior mean halo size is $5.4 \pm 1.4$  kpc, also in agreement with our
earlier estimated range $z_h=4-6$ kpc \citep{SM2001a}, although our best-fit 
value of $z_h = 3.9$ kpc is somewhat lower, due to the degeneracy 
between $D_0$ and $z_h$. 
However, the well-defined posterior intervals produced by the present study are 
significantly more valuable than just the best fit values themselves as they 
provide an estimate of associated 
theoretical uncertainties and may point to a potential inconsistency 
between different types of data.  

\begin{figure*}
\centerline{
\includegraphics[width=0.5\linewidth]{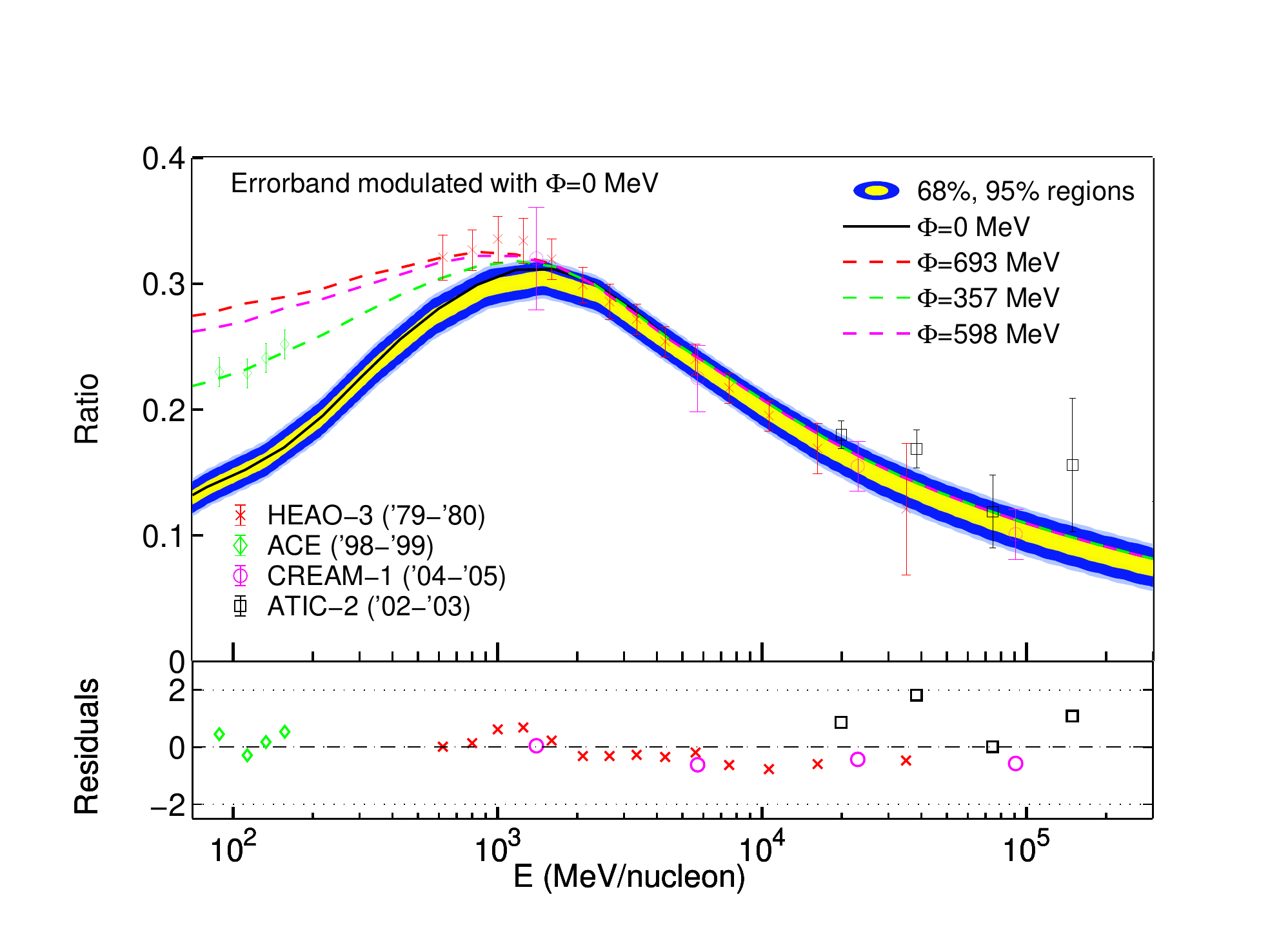}
\includegraphics[width=0.5\linewidth]{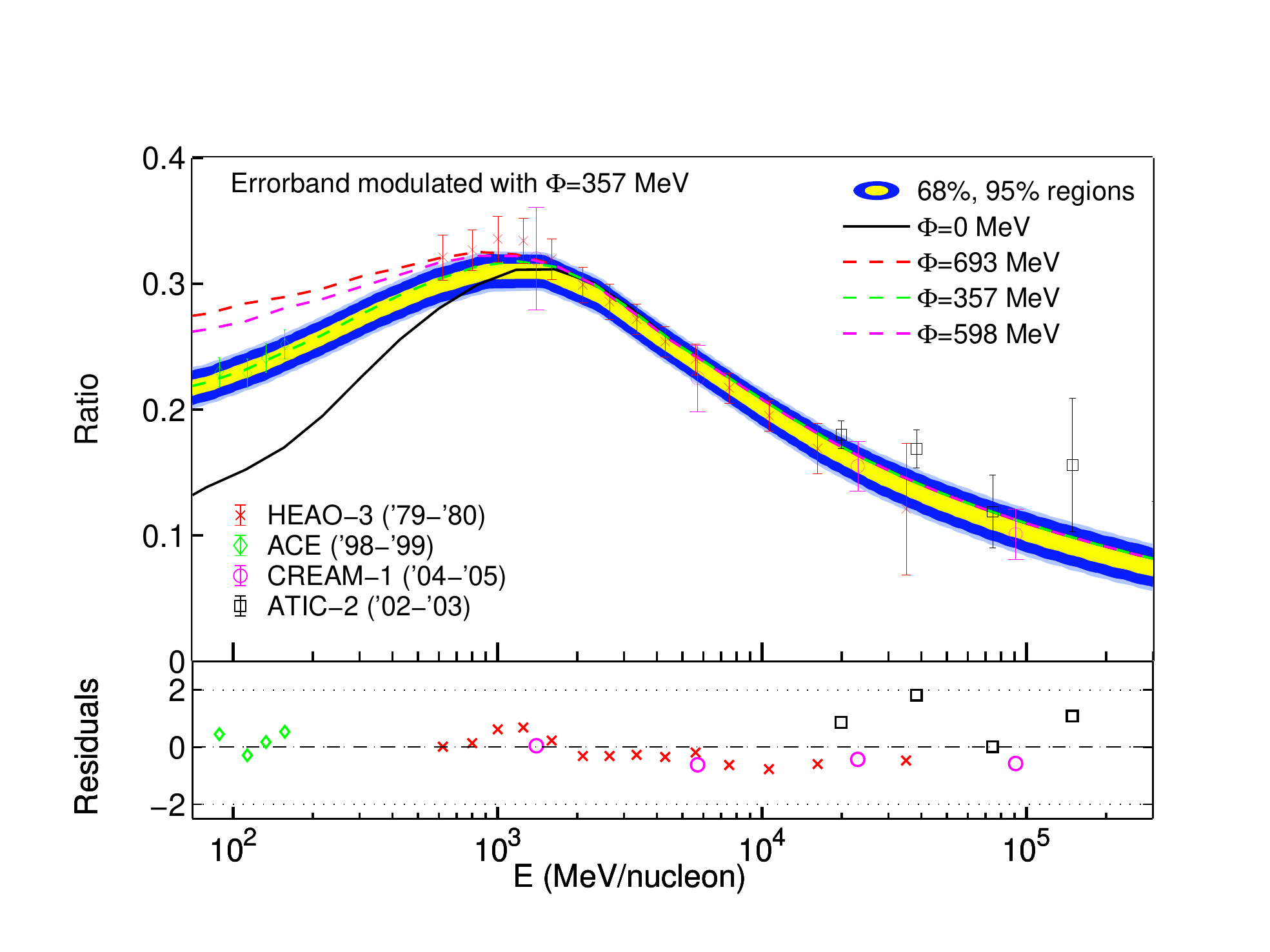}
}
\centerline{
\includegraphics[width=0.5\linewidth]{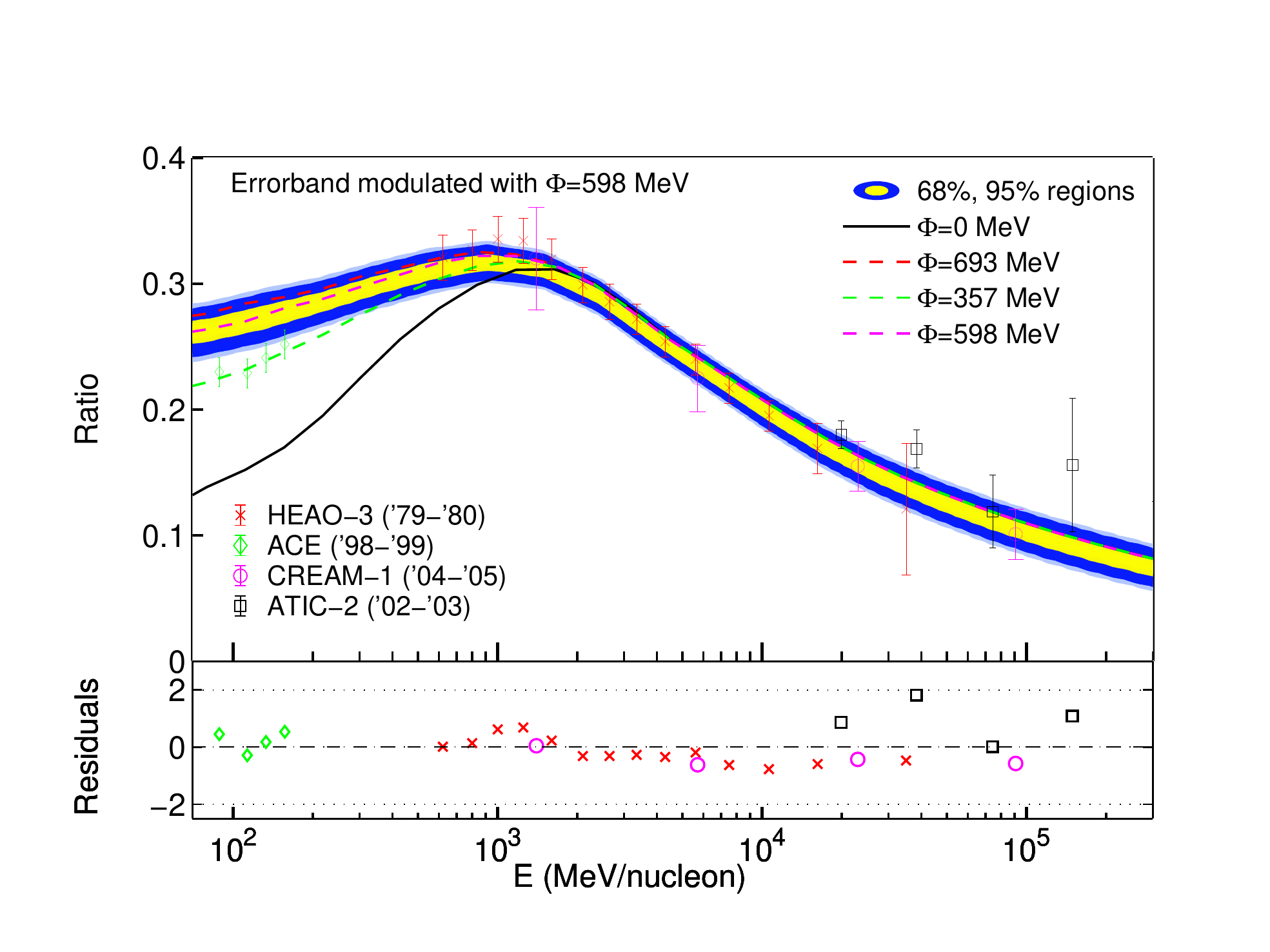}
\includegraphics[width=0.5\linewidth]{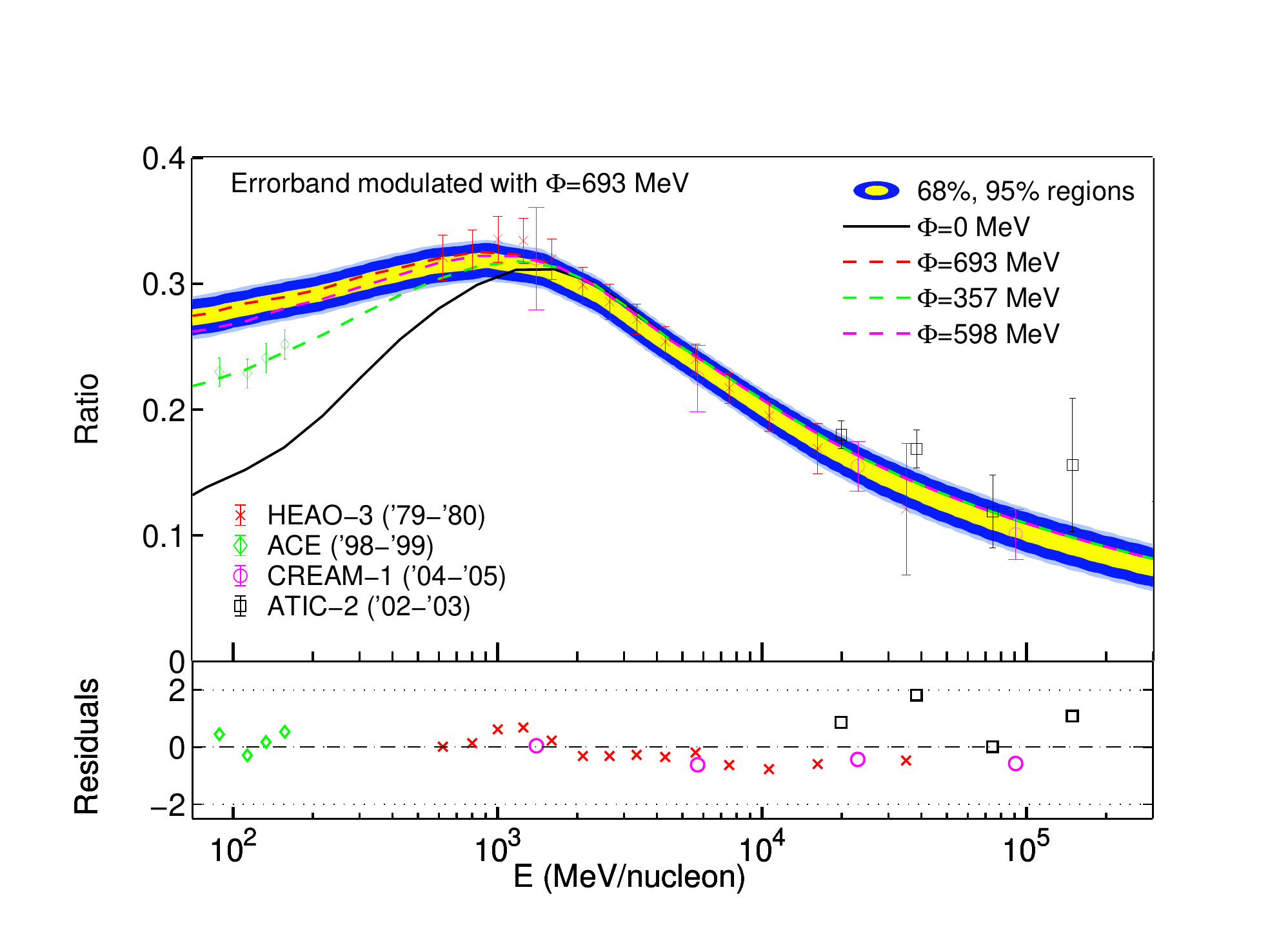}
}
\caption{B/C ratio for our best fit parameters (dashed curves). 
Each of the dashed curves has been modulated with the best-fit potential 
from our global fits, with value given in the legend. 
We also plot the fitted datasets, each with error bars enlarged by the 
best-fit value of our scaling parameters, $\tau$, as given 
in \protect{Table~\ref{tab:params_constraints}}. 
Color coding of each data set matches the color of the best-fit 
modulated curve with which it should be compared: ACE \citep[solar minimum,][]{George2009} with $\Phi=357$ MV, 
CREAM \citep{Ahn2008} with $\Phi=598$ MV, HEAO-3 \citep{Engelmann1990} with $\Phi=693$ MV and ATIC 
\citep{Panov2008} with $\Phi=0$ MV (no modulation), see the description of the data in the text. 
The yellow/blue error bands delimit regions of 68\% and 95\% 
probability, and are modulated according to the potential given in 
each panel. 
The bottom of each panel shows the residuals of our best-fit model, 
defined in \protect{eq.~\eqref{eq:residuals}}. 
\label{BC}}
\end{figure*}

\subsection{Quality of Best-Fit model}

We now assess the quality of our best-fit model. 
Define the $\chi^2$ as 

\be
\chi^2 \equiv \sum_{j=1}^5 \sum_{i=1}^{N_j}  \frac{\left(\Phi_X(E_i, \params, \phi) - \hat{\Phi}_X^{ij} \right)^2}{\sigma_{ij}^2/\tau_j},
\ee

\noindent
i.e., we compute the $\chi^2$ using the rescaled error bars for the data 
points (notice that  $\chi^2 \neq -2 \log P(\data|\params, \phi, \tau)$, 
i.e.~the $\chi^2$ is not minus twice 
the log-likelihood because of the pre-factor containing $\tau$ 
appearing in Eq.~[\ref{eq:Gauss_like_rescaled}]).
There are $N = 76$ total data points and $M=16$ fitted parameters, 
including both the modulation 
and the error rescaling parameters. 
Therefore the number of degrees of freedom (dof) is 60, and for the 
best-fit model we 
find $\chi^2 = 69.3$, which leads to a reduced 
chi-squared $\chi^2/\text{dof} = 68/60 = 1.15$.
This is not surprising, since by construction the error bar rescaling 
parameters, $\tau$, 
are adjusted dynamically during the global fit to achieve this. 
A more detailed breakdown of the contribution to the total $\chi^2$ by data set 
is given in Table~\ref{tab:residuals}. 

\begin{figure*}
\centerline{
\includegraphics[width=0.5\linewidth]{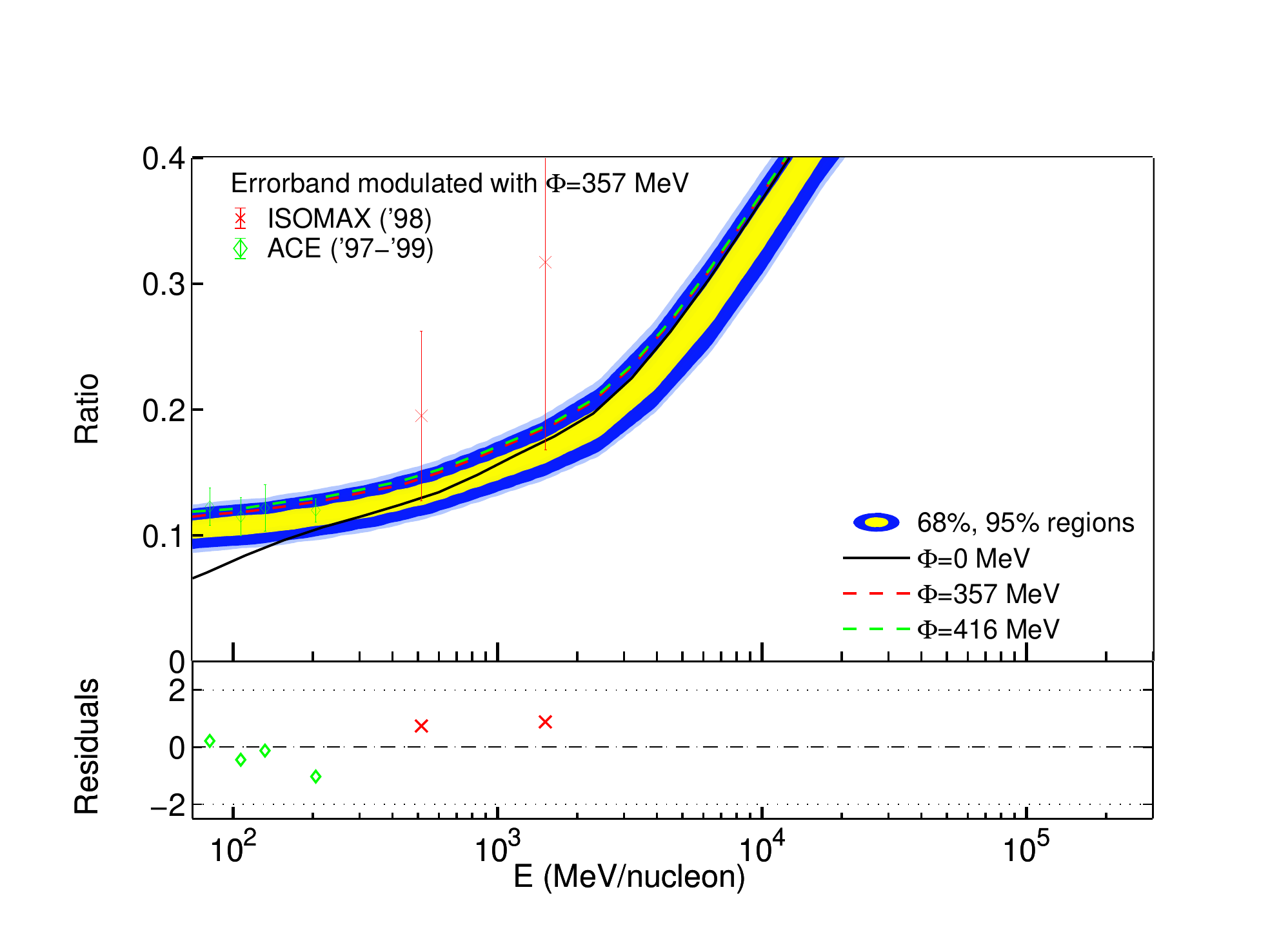}
\includegraphics[width=0.5\linewidth]{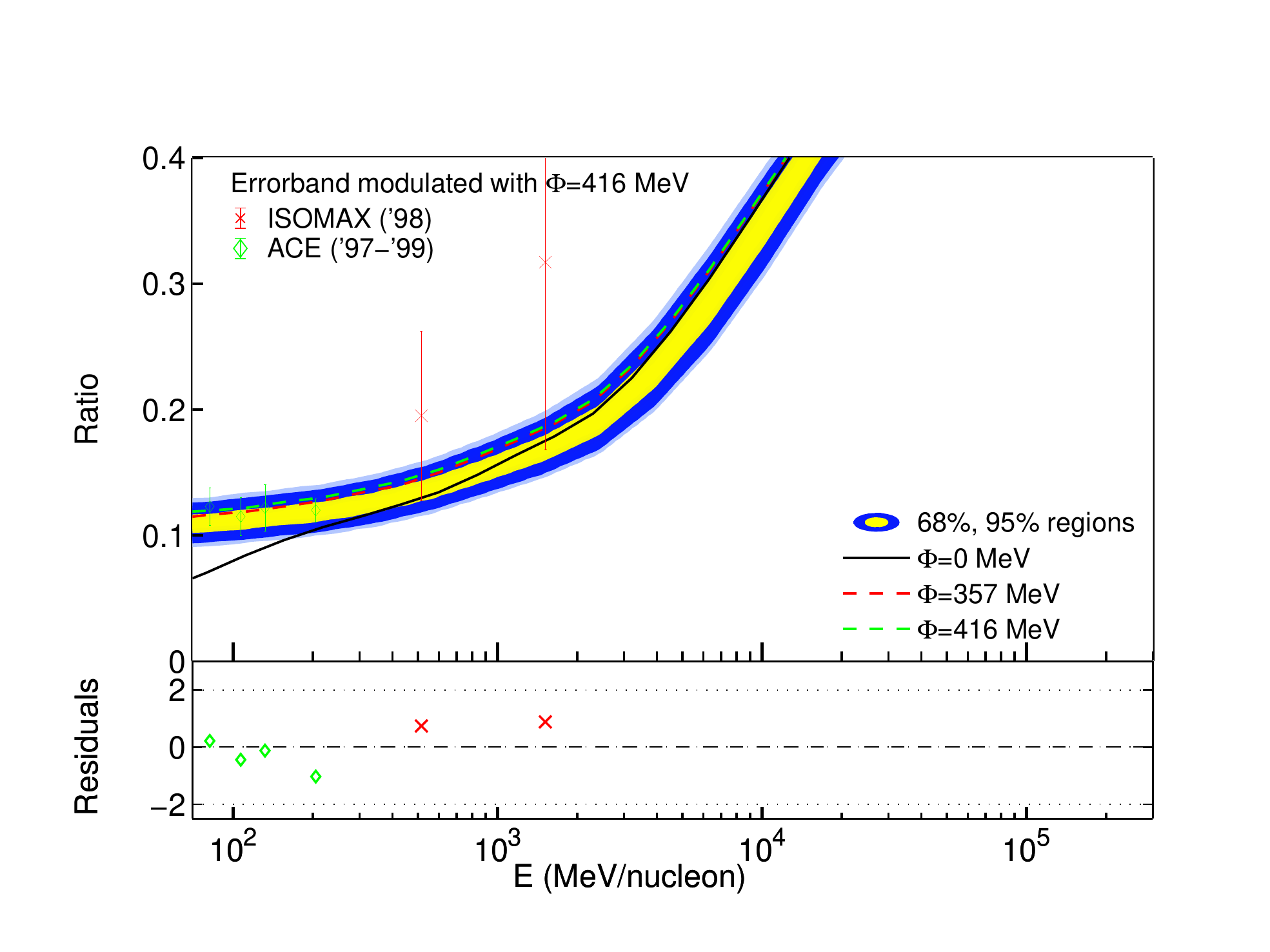}
}
\caption{$^{10}$Be/$^9$Be ratio for our best fit parameters, including 
error bands, as in \protect{Fig.~\ref{BC}}. Color coding of each data set matches the color of the best-fit 
modulated curve with which it should be compared: ACE \citep{Yanasak2001} with $\Phi=357$ MV, 
ISOMAX \citep{Hams2004} with $\Phi=416$ MV.
We also plot the unmodulated ratio for comparison.  
\label{Be}}
\end{figure*}

\begin{figure*}
\centerline{
\includegraphics[width=0.5\linewidth]{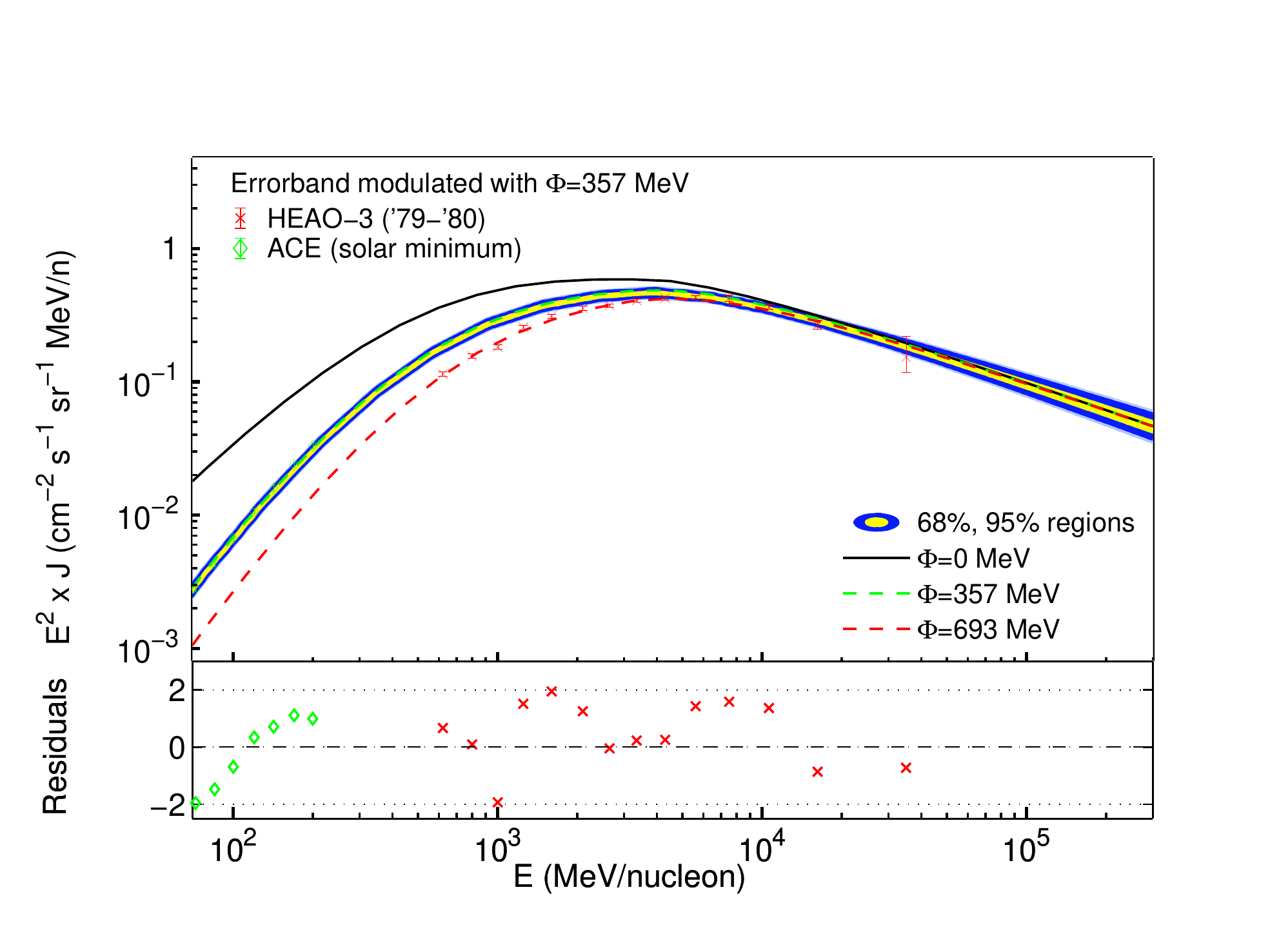}
\includegraphics[width=0.5\linewidth]{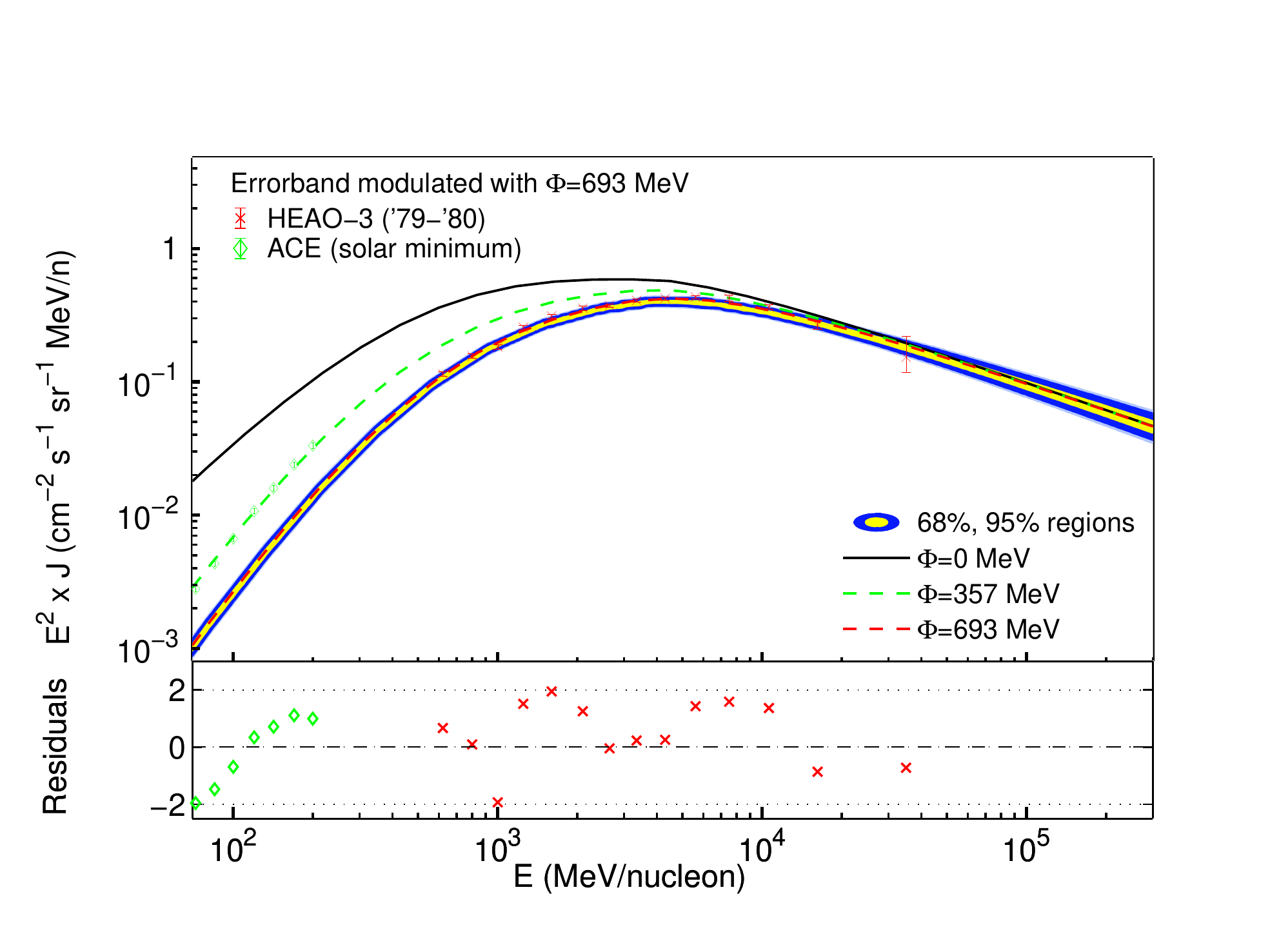}
}
\centerline{
\includegraphics[width=0.5\linewidth]{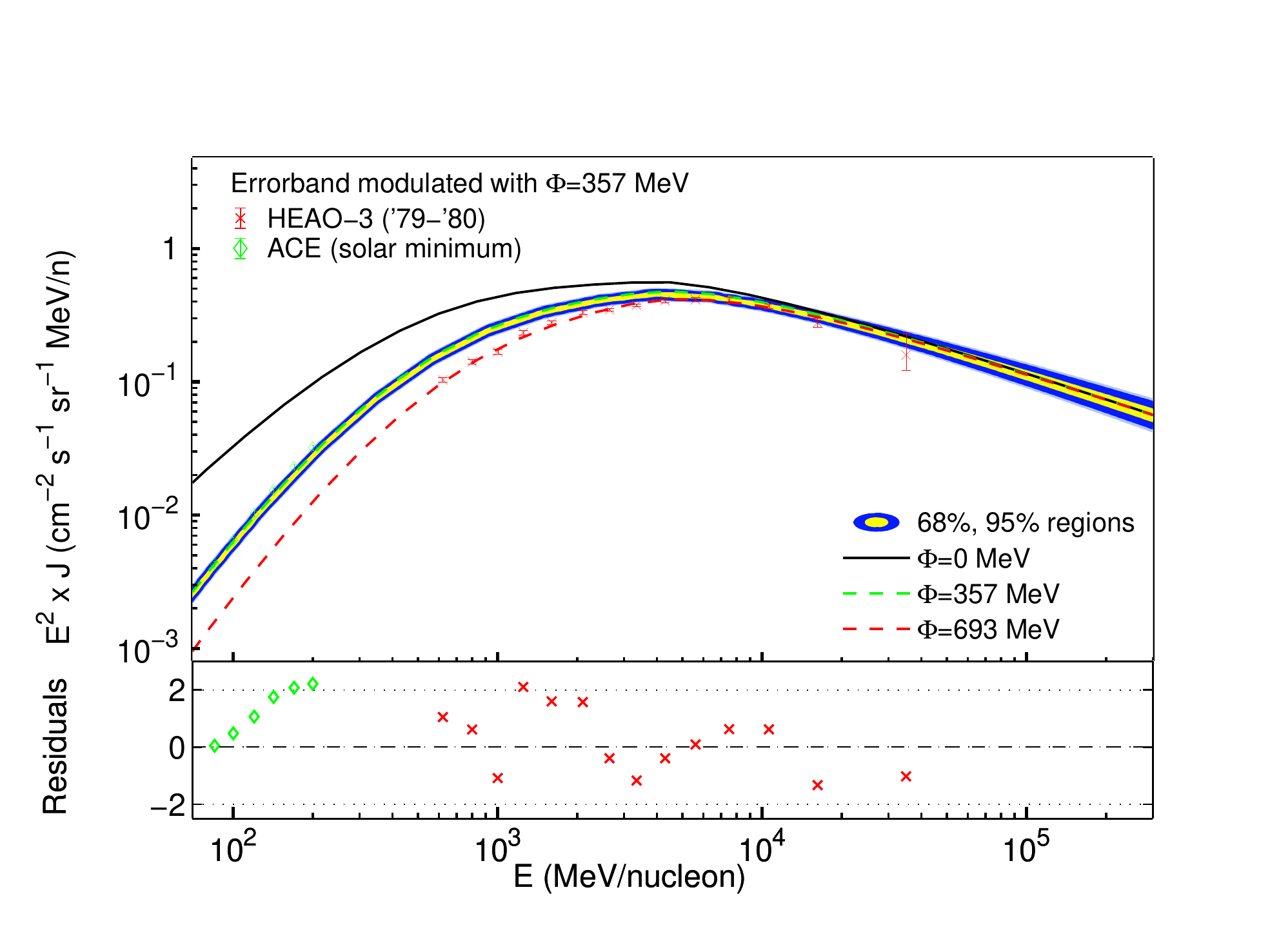}
\includegraphics[width=0.5\linewidth]{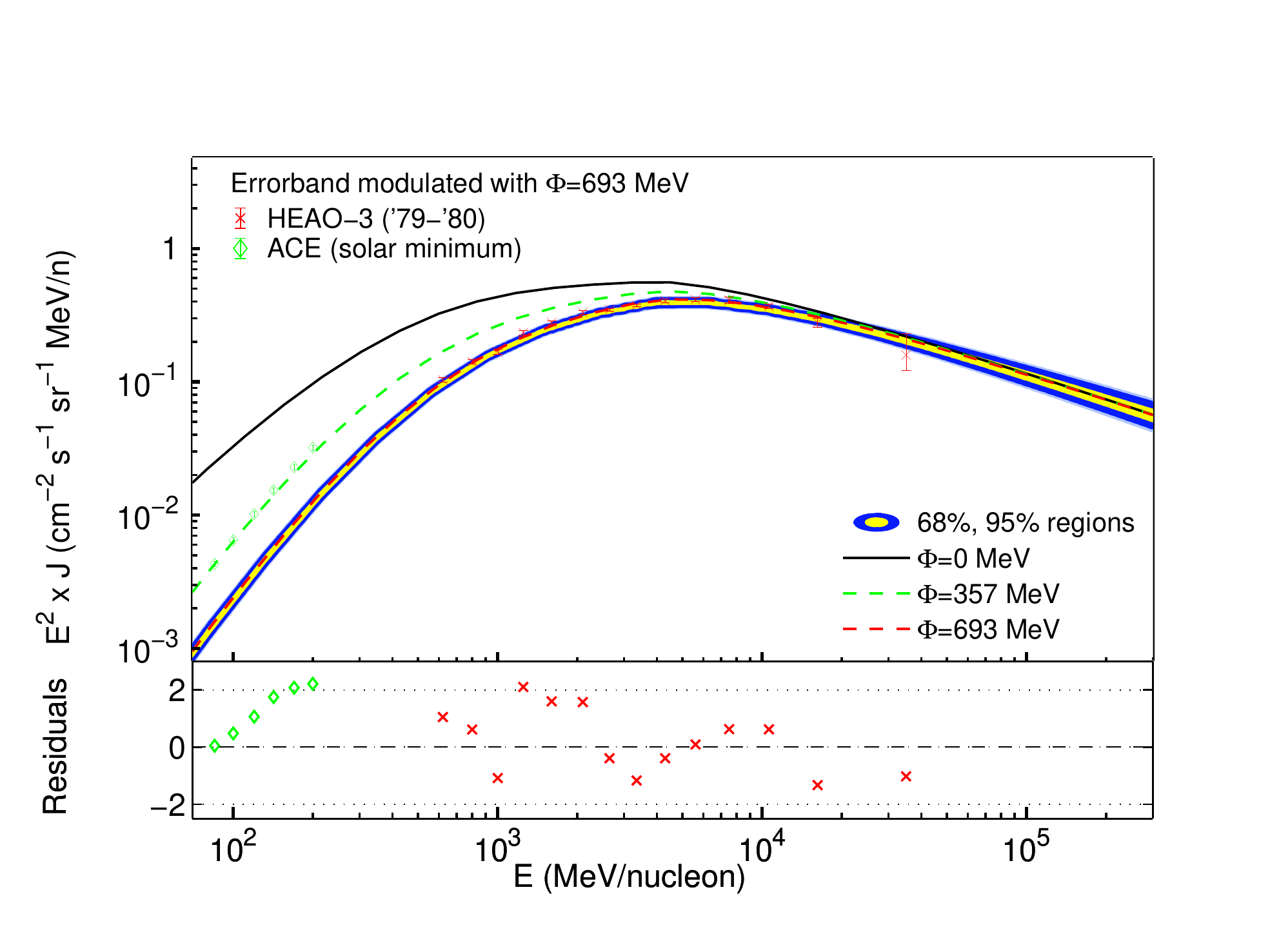}
}
\caption{Carbon (top panels) and oxygen (bottom panels) spectra for our 
best fit parameters, including error bands and best-fit model residuals, 
as in \protect{Fig.~\ref{BC}}. Color coding of each data set matches the color of the best-fit 
modulated curve with which it should be compared: ACE \citep[solar minimum,][]{George2009} with $\Phi=357$ MV, 
HEAO-3 \citep{Engelmann1990} with $\Phi=693$ MV.
We also plot the unmodulated spectrum for comparison. 
\label{CO}}
\end{figure*}

The predictions for the fitted CR spectra of the best-fit model 
parameters are shown in Figures~\ref{BC}-\ref{CO}, including 
an error band delimiting the 68\% and 95\% probability regions.
The species shown are B/C and $^{10}$Be/$^9$Be ratios, and the spectra 
of carbon and oxygen. 
In each plot, we show the spectrum modulated with the potential 
corresponding to our best-fit 
parameters from our global fits for each of the data sets employed. 
We also show the datasets, each with error bars enlarged by the best-fit value 
of our scaling parameters, $\tau$, as 
given in \protect{Table~\ref{tab:params_constraints}}. 
The yellow/blue band delimits regions of 68\% and 95\% probability, and is 
modulated 
according to the potential given in the each panel. 
We emphasize that the power of our statistical technique is such 
that we can, for the first time, 
provide not only a best fit model but also an error band with a 
well-defined statistical meaning. 

In order to better visualize the comparison of our best-fit model to 
the fitted data, we 
plot in the bottom part of each panel the best-fit residuals i.e., the 
difference between 
data and best-fit model, divided by the experimental 
error bar (enlarged by the correct error scaling parameter): 

\be \label{eq:residuals}
{\mathcal R}_{ij} = \frac{ \hat{\Phi}_X^{ij} - \Phi_X(E_i, \params, \phi)}{\sigma_{ij}/\sqrt{\tau_j}}.
\ee

\begin{deluxetable*}{cccccc}[t]
\tabletypesize{\footnotesize}
\tablecaption{\label{tab:residuals} 
Breakdown of contributions to the total $\chi^2$ of our best fit by data set
}
\tablecolumns{5}
\tablewidth{0pt}
\tablehead{
CR & Data sets & Data points, $n$ & $\chi^2$ & $\chi^2/n$}
\startdata
Oxygen & HEAO-3, ACE & 20 & 28.9 & 1.44  \\ 
Carbon & HEAO-3, ACE & 21 & 29.5 & 1.40  \\ 
B/C & HEAO-3, ACE, ATIC-2, CREAM & 29 & 8.9 & 0.30  \\
$^{10}$Be/$^9$Be & ACE, ISOMAX & 6 & 2.0 & 0.33 \\
\hline
All & All & 76 & 69.3 & $\chi^2/\text{dof} = 1.15$
\enddata
\end{deluxetable*}

\noindent
Visual inspection of the residuals for the B/C and the $^{10}$Be/$^9$Be 
ratios (see Figures~\ref{BC} and \ref{Be}) shows that our best-fit 
model gives 
an excellent fit to those data, with the distribution of the residuals 
approximately symmetric around 0. 
This indicates that there is no systematic bias of our best-fit. 
The contribution to the overall $\chi^2$ from those data sets 
is, if anything, smaller 
than would be expected statistically: Table~\ref{tab:residuals} indicates 
that each datum contributes about $\sim 0.3$ units to the $\chi^2$. 
This could point to a degree of overfitting, or to our error bar rescaling 
parameters being too small. 
However, the origin of this slight overfitting becomes clear when one considers 
the oxygen and carbon spectra, and their residuals (Figures~\ref{CO}). 
Residuals here are significantly larger, especially at low 
energies, $E  < 3$ GeV, 
and the average contribution to the total $\chi^2$ by each datum is 
much larger, of order $\sim 1.4$, see Table~\ref{tab:residuals}. 
Therefore, the error bars on carbon and oxygen seem to require enlargement in 
order for our model to provide a good fit. 
Notice from the shape of the residuals in Fig.~\ref{CO} that there is no 
indication of systematic bias in our fit, i.e., the enlargement of the 
error bars does not come about because the model cannot reproduce the 
data, rather, because the data themselves seem to show an amount of scatter 
that is incompatible with a smooth spectrum (unless the error bars are 
enlarged sufficiently). 

As a consequence, we can conclude that it is the carbon and oxygen 
spectra that are 
driving the value of the error bars rescaling parameters for HEAO-3 and ACE.
Therefore, the error bars are correspondingly enlarged in 
the B/C and $^{10}$Be/$^9$Be 
spectra, since we are using one single error bar rescaling parameter 
for each experiment. 
This results in much smaller residuals for the latter spectra. 
Introducing a larger number of error bar rescaling parameters, one for 
each experiment and for each CR species, 
and fitting them independently could resolve this issue. 
Then, the rescaling will be less important for B/C and $^{10}$Be/$^9$Be, 
leading to tighter constraints from those data sets.

\begin{figure*}[t!]
\centerline{
\includegraphics[width=0.5\linewidth]{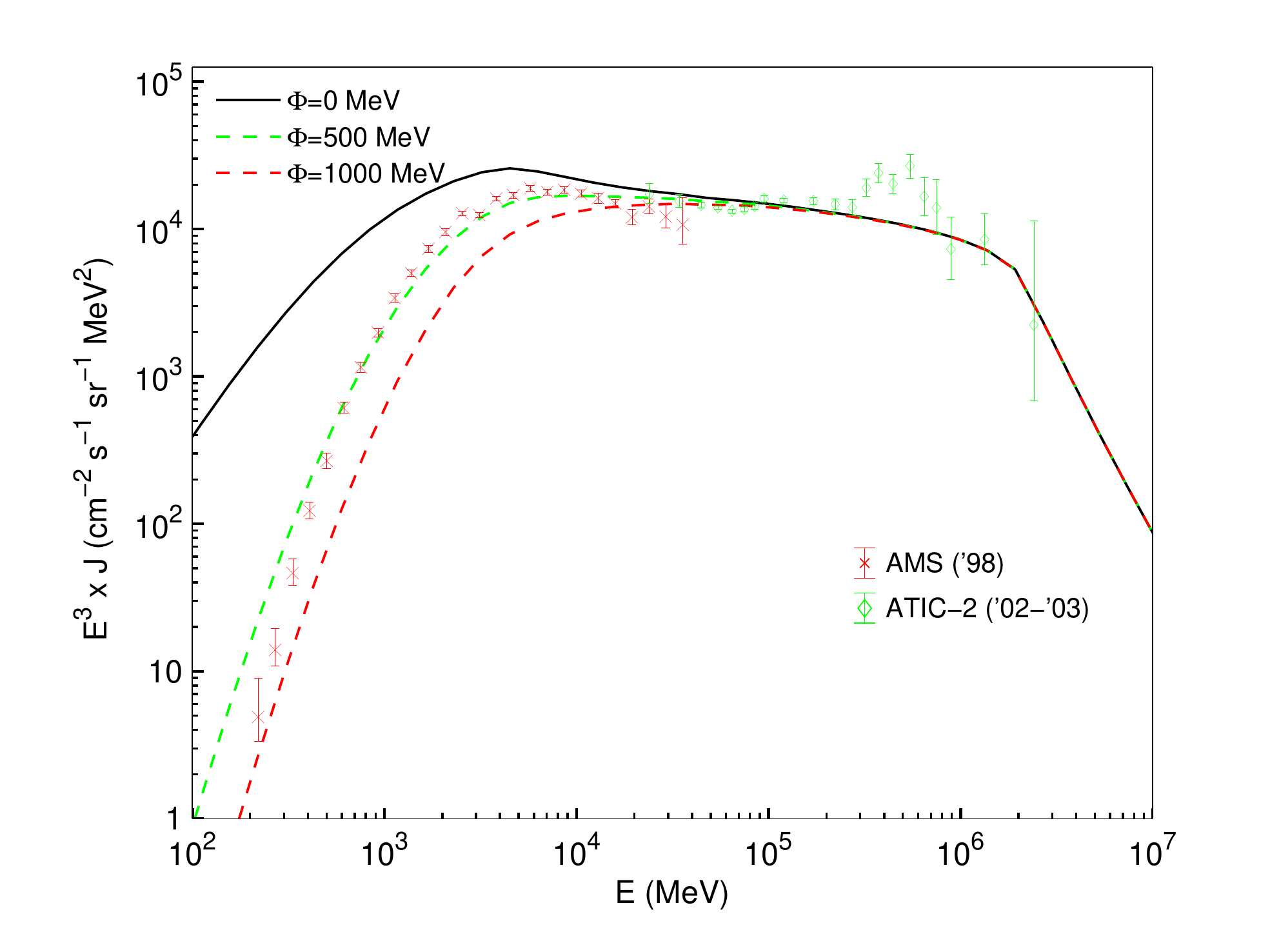}
\includegraphics[width=0.5\linewidth]{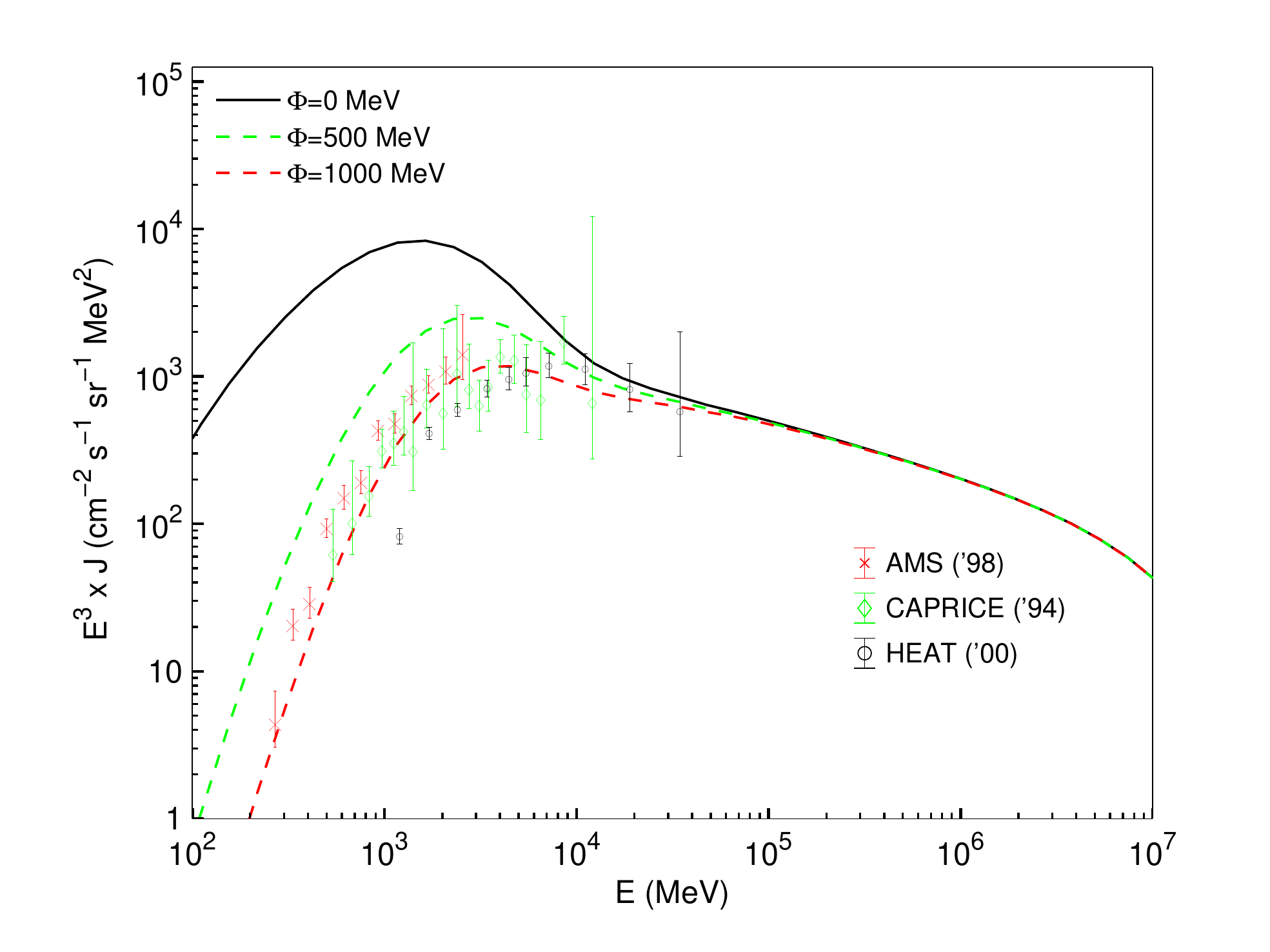}
}
\caption{Electrons (left panel) and positrons (right panel) spectra for 
our best fit parameters, with three choices of modulation. 
As for the antiproton spectrum and diffuse \gray{s}, electrons and positrons spectra have 
not been used in the fit, therefore the lines should be interpreted as 
predictions from our model. 
We show experimental data on each quantity as 
well: electrons -- AMS-01~\citep{AMS2000}, ATIC-2~\citep{ATIC2008}, 
HESS~\citep{HESS2008,HESS2009}, positrons -- AMS-01~\citep{AMS2007}, 
CAPRICE~\citep{CAPRICE2000}, HEAT~\citep{HEAT2004}. The dates in the legend for the data sets give the years when the corresponding data were collected.
 \label{ele}}
\end{figure*}

\begin{figure*}[t!]
\centerline{
\includegraphics[width=0.5\linewidth]{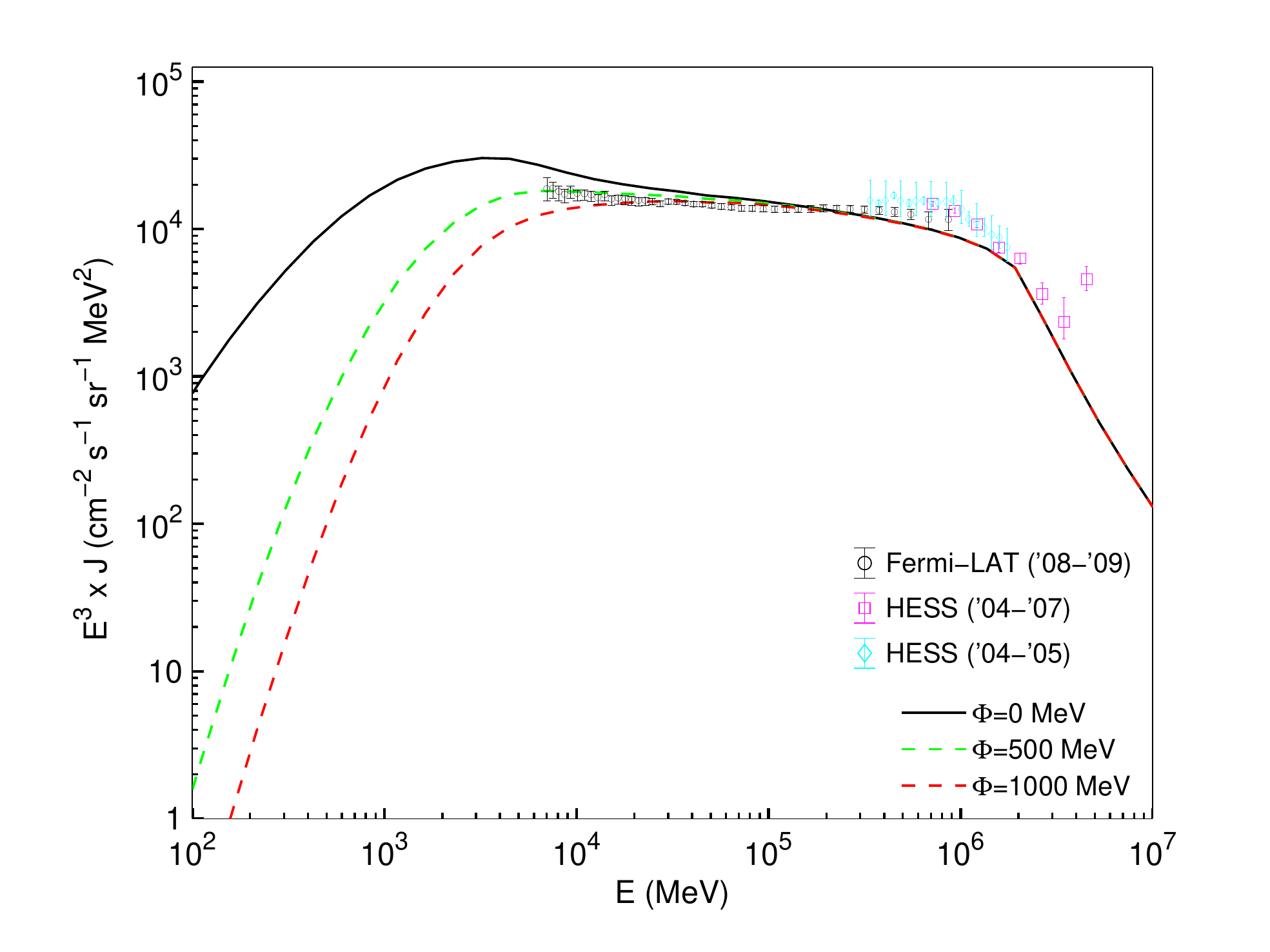}
\includegraphics[width=0.5\linewidth]{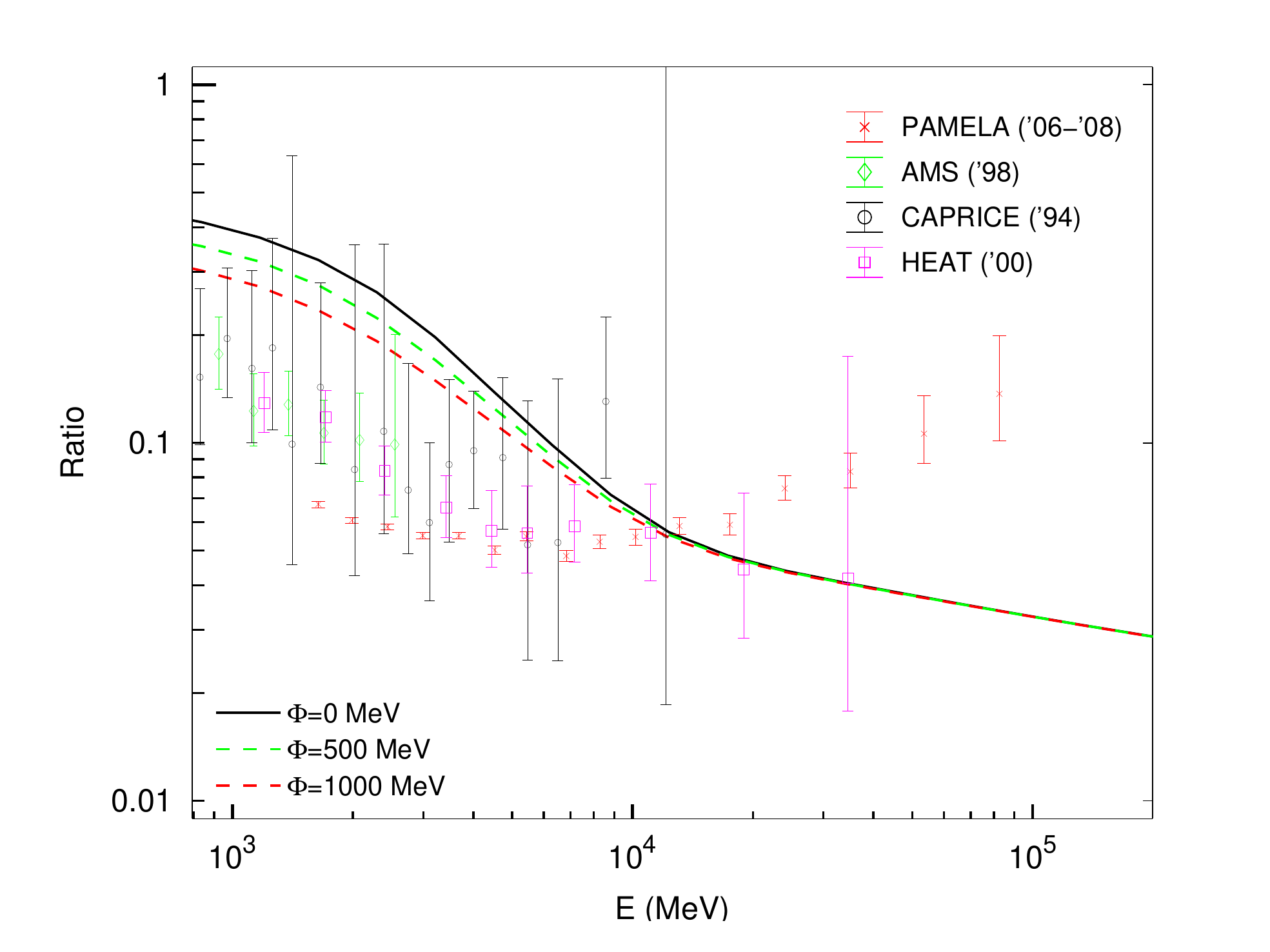}
}
\caption{Left panel: total leptons (positrons plus electrons) spectrum for our best-fit CR parameters, with three choices of modulation potential. We also show the data from {\it Fermi}-LAT~\citep[a sum of electrons and positrons,][]{Abdo2009electrons,Abdo2010electrons}.  Right panel: corresponding positron fraction, with the same three 
choices of solar modulation potential. 
We show experimental data from PAMELA \citep{Adriani2009}, 
AMS01 \citep{AMS2007}, CAPRICE \citep{CAPRICE2000}, 
and HEAT \citep{HEAT2004}. These 
data sets have {\it not} been fitted in the analysis. 
\label{pos_ratio}}
\end{figure*}

\pagebreak[4]
\section{Discussion} \label{discussion}

\subsection{Implications for Antimatter and \gray{s}} 

\begin{figure}
\centerline{
\includegraphics[width=1\linewidth]{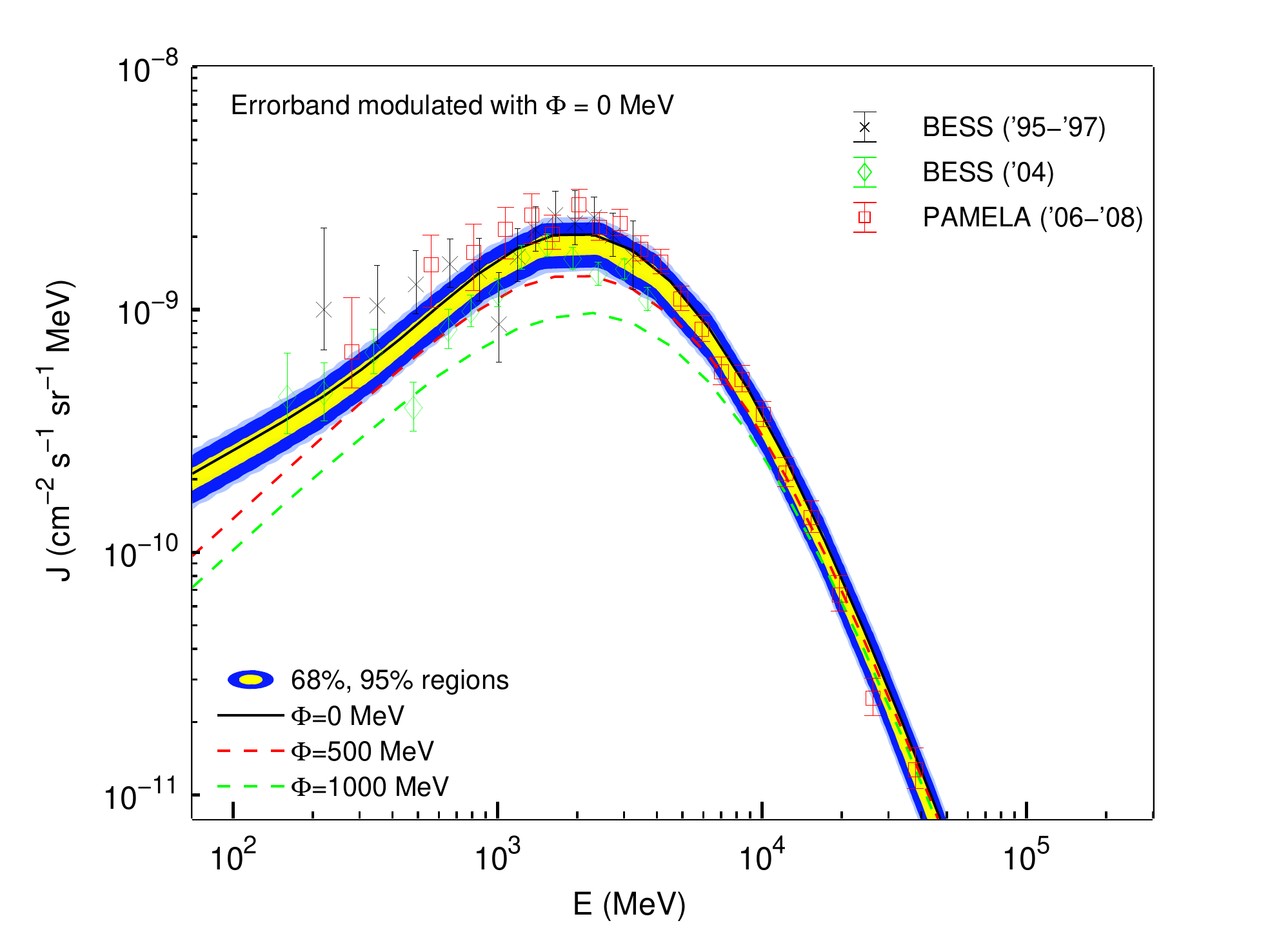}
}
\caption{Antiproton spectrum for our best-fit CR parameters, with 
three different representative solar modulation potentials together with 
recent data. 
Note that the error band has not been modulated. 
The modulated curve for $\Phi$=500 MV is most appropriate 
for the BESS 1995--1997 flight \citep{BESS2000} and PAMELA current solar 
minimum data \citep{Adriani2010}, while the BESS-Polar flight of 2004 \citep{BESS2008} corresponds to the higher level of solar activity. 
The data shown here have {\em not} been fitted. 
 \label{pbar}}
\end{figure}

We also use our best-fit model to calculate secondary antiprotons, 
electrons, positrons, and diffuse \gray{s} 
which were not fitted, but provide a useful consistency check. 
For this calculation the spectra of CR protons and He were adjusted to 
the BESS data \citep{BESS2000protons}. 
Secondary antiprotons were calculated using the same formalism 
as in \citet{Moskalenko2002}.
Calculation of secondary electrons, positrons, and diffuse \gray{s} is 
described in Sec.~\ref{galprop}.
The spectra of secondaries are shown in Figs.~\ref{ele}-\ref{pbar}.
The primary electron injection spectrum is based on fitting to pre-Fermi electron data \citep[conventional model,][]{SMR2004,Ptuskin2006},
 and is parameterized as a broken power law with indices 1.6/2.5 below/above 4 GeV and a steepening (index 5)
above 2 TeV, and normalized to the Fermi data at 25 GeV \citep{Abdo2010electrons}.
The plot for the CR electron and positron spectrum is shown separately in Fig.~\ref{ele} and compared to relevant measurements, 
while the plot for total leptons (electrons plus positrons) is displayed in the left panel of Fig.~\ref{pos_ratio}. 
Even though the total electron and diffuse emission data 
were not fitted, they agree well with our best-fit model predictions. 
The positron fraction, shown in the right panel of Fig.~\ref{pos_ratio}, does not agree with the 
PAMELA data \citep{Adriani2009}, but this was expected since secondary 
positron production in the general ISM is not capable of producing an 
abundance that rises with energy.

Antiprotons, shown in Fig.~\ref{pbar}, also not fitted, present an interesting example where 
the intensity at a few GeV is significantly underpredicted by the 
reacceleration models.
As has been already shown \citep{Moskalenko2002,Moskalenko2003}, the 
antiproton flux measurements by BESS taken during the last solar 
minimum, 1995--1997 \citep{BESS2000}, are inconsistent with 
reacceleration models at the 40\% level at about 2 GeV, while the 
stated measurement uncertainties in this energy range are 20\%. 
The reacceleration models considered
are conventional models, based on local CR 
measurements, Kolmogorov diffusion, and uniform CR source spectra 
throughout the Galaxy.
Models without reacceleration that can reproduce the 
antiproton flux, however, fall short of explaining the 
low-energy decrease in the secondary/primary nuclei ratio.
To be consistent with both, the introduction of breaks in the 
diffusion coefficient and the injection spectrum is required, which 
may suggest new phenomena in particle acceleration and propagation. 
An inclusion of a local primary component at low energies, 
possibly associated with the Local Bubble, can reconcile 
the data \citep{Moskalenko2003}.

Figure~\ref{pbar} shows that the reacceleration model underproduces 
antiprotons in the GeV range by $\sim$30\% also compared to the 
PAMELA data taken during the current solar minimum \citep{Adriani2010}. 
On the other hand, high-energy PAMELA data agree well with the 
predictions, which may suggest that the excess over the model predictions at low energies 
can be associated with the solar modulation.
Since reacceleration is also most important below a few GeV, at 
present it does not appear possible to distinguish these effects.
However, a more systematic analysis which includes evaluation of 
other propagation models may help and will be carried out in a future work.
 


\subsection{Comparison with other Analyses}

Most other analyses have used analytical propagation models, in particular
\citet{Donato2002} and \citet{Maurin2002}. 
The most recent reported results from this approach are in
\citet{Maurin2010} and \citet{Putze2010},
which used $\chi^2$ and MCMC techniques to analyze a wide 
range of semi-analytical models. 
The nearest case to ours is their diffusion/reacceleration model where 
they found  $\delta = 0.23-0.24$, $\valf = 70-80$ km s$^{-1}$ 
and $z_h = 4-6$ kpc. 
Better fits are found with convection included as well, with a 
smaller $\valf$, but this 
requires a very large $\delta = 0.8-0.9$. 
In contrast, we find a more plausible range of values for the 
diffusion/reaccleration model (however we do not consider convection here). 

We also note that they assume an injection spectrum that is a 
single power-law in rigidity, which also includes a factor of $\beta^{\eta_S}$, $\sim$$\beta^{\eta_S}\rho^{-\nu}$ \citep{Putze2010a}.
In the case of proton and He injection spectra in the reacceleration model (their Model II)
they use $\eta_S\approx 1$.  This is equivalent to our break in the injection spectrum provided the diffusion 
coefficient has a form Eq.~(\ref{Dxx}).
For heavier nuclei, C to Fe,
their injection spectrum has $\eta_S\approx 2$, which compensates for the large values of the $\valf$ in their fits.
Therefore, their best fit parameters (e.g., $\valf$) are not equivalent to ours and are dependent on a particular 
choice of the injection spectrum. Their conclusion that they do not need a break in the
injection spectrum is thus significantly overstated as they need an \emph{ad hoc} factor $\beta^{\eta_S}$ with
index ${\eta_S}$ different for different groups of nuclei.


\section{Conclusions}

For the first time, 
we have shown that a full Bayesian analysis is possible using both 
MCMC and nested sampling despite the heavy computational demands of a 
numerical propagation code. 
Furthermore, our analysis also returns the value of the Bayesian 
evidence, which could be used to rank different propagation models 
in terms of their performance in explaining the data. 
While we have not investigated this possibility here, we leave this 
topic for a dedicated discussion in a forthcoming publication. 

\begin{figure}[tb!]
\centerline{
\includegraphics[width=0.9\linewidth]{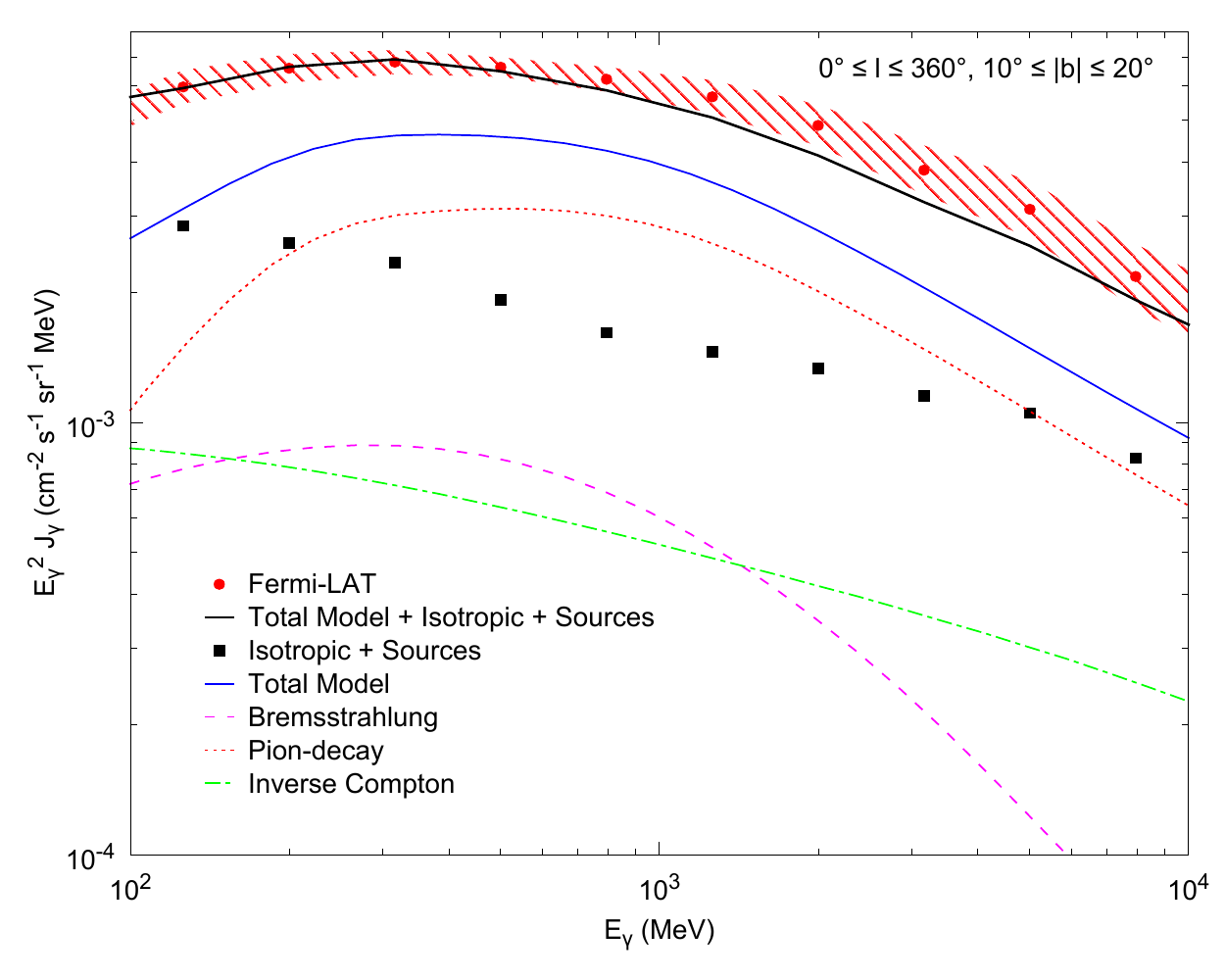}
}
\caption{Diffuse \gray{} spectra for $10^\circ \leq |b| \leq 20^\circ$ for 
our best-fit CR parameters compared with {\it Fermi}-LAT data for the same
region of sky.
The {\it Fermi}-LAT data along with the unidentified background 
and source components 
are taken from \citet{Abdo2009diffuse}. The red hatched area represents 
the systematic uncertainty in the spectrum of the diffuse emission, as given
in \citet{Abdo2009diffuse}.
Note that our best-fit model corresponds closely to that used in 
\citet{Abdo2009diffuse} to derive the unidentified background 
and source components.
Therefore, the addition of our model to these components is valid.
Note that these data have {\it not} been fitted.
\label{gammas}}
\end{figure}

The present study provides not just best-fit values for the 
propagation parameters but, more importantly, associated uncertainties 
which fully account for correlations among parameters, as well as 
for experimental and theoretical uncertainties. 
Such error estimates have been fully propagated to the predicted CRs 
spectra from the model, thus providing an estimate of the residual 
uncertainty on the predictions (after fitting), which can be used to 
assess, e.g., potential 
inconsistencies between different types of data or for model selection. 
An important conclusion is that the parameter ranges derived in this study
are consistent with previous, less systematic analyses.

A valuable test made possible by our technique is the consistency 
check of the calculated spectra of 
other CR species, antiprotons, electrons, 
and positrons with data. 
Total electrons agree with the {\it Fermi}-LAT spectrum within the systematic uncertainties, see Fig.~\ref{gammas}.
The calculated spectrum of the Galactic diffuse emission for the 
mid-latitude range is 
also in a good agreement with the observations by the {\it Fermi}-LAT.
The antiprotons are under-predicted, which is a general feature of reacceleration models as we have shown before.
The positron fraction is inconsistent with a purely secondary origin of positrons above 1 GeV, confirming the 
excess above 10 GeV attributed to sources of primary positrons. 
On the other hand, the predicted positron {\it spectrum} is consistent with earlier experiments given their large error bars. We await publication of
PAMELA positron data for further model comparison.

Because of the pioneering nature of our approach, we have 
concentrated on just one particular type of 
model with reacceleration and a power-law diffusion coefficient. 
Other propagation modes, e.g., convection, are not considered here. 
Future work will consider these, together with constraints from antiprotons, 
positrons, and \gray{s}.

\acknowledgements  
R.~T. would like to thank the EU FP6 Marie Curie Research and Training Network
``UniverseNet'' (MRTN-CT-2006-035863) for partial support and 
the Institut d'Astrophysique de Paris for hospitality. 
I.~V.~M. acknowledges support from NASA~Grant~No.~NNX09AC15G.
T.~A.~P. acknowledges support from NASA~Grant~No.~NNX10AE78G. The use of Imperial College High Perfomance Computing cluster is gratefully acknowledged. 

\appendix
\section{Supplementary Material}

The Supplementary Material accompanying this paper contains the following files, which can be used to reproduce the results presented in the paper. 

\begin{itemize}
\item {\em Galprop\_chains.tar.gz}: 
Upon unpacking, this archive contains the 10 MCMC chains used in the paper, and one .info file detailing what information is contained in the chains.
\item The folder {\em BestFitSpectra} contains datafiles with the best-fit spectra from our paper. The files and their contents are described in an accompanying {\em README} file.
\item Galprop definition files are supplied for the best-fit parameter values as well as for the posterior mean values (as per Table~\ref{tab:params_constraints}).

\end{itemize}

\bibliographystyle{apj}
\bibliography{imos,strong,FermiDiffusePaper2,extragalactic,GalpropBayes}{}



\end{document}